\documentclass[pdflatex,sn-mathphys-num]{sn-jnl}

\usepackage{graphicx}%
\usepackage{multirow}%
\usepackage{amsmath,amssymb,amsfonts}%
\usepackage{amsthm}%
\usepackage{mathrsfs}%
\usepackage[title]{appendix}%
\usepackage{xcolor}%
\usepackage{textcomp}%
\usepackage{manyfoot}%
\usepackage{booktabs}%
\usepackage{algorithm}%
\usepackage{algorithmicx}%
\usepackage{algpseudocode}%
\usepackage{listings}%
\usepackage{subcaption}
\usepackage{array}
\newcolumntype{P}[1]{>{\raggedright\arraybackslash}p{#1}}

\theoremstyle{thmstyleone}%
\theoremstyle{thmstyletwo}%

\theoremstyle{thmstylethree}%

\begin{document}

\title[Article Title]{Ferroelectric FET-based Logic-in-Memory Encoder for Hyperdimensional Computing}


\author[1]{\fnm{Arka} \sur{Chakraborty}}\email{arkac21@iitk.ac.in}

\author[2]{\fnm{Franz} \sur{Müller}}

\author*[2,3]{\fnm{Thomas} \sur{Kämpfe}}\email{thomas.kaempfe@ipms.fraunhofer.de}

\author[1]{\fnm{Shubham} \sur{Sahay}}\email{ssahay@iitk.ac.in}

\affil[1]{\orgdiv{Department of Electrical Engineering}, \orgname{Indian Institute of Technology Kanpur}, \orgaddress{\street{Kalyanpur}, \city{Kanpur}, \postcode{208016}, \state{Uttar Pradesh}, \country{India}}}

\affil[2]{\orgdiv{Center Nanoelectronic Technologies}, \orgname{Fraunhofer IPMS}, \orgaddress{\street{An der Bartlake 5}, \city{Dresden}, \postcode{01109}, \country{Germany}}}

\affil[3]{\orgdiv{Institute of CMOS Design}, \orgname{TU Braunschweig}, \orgaddress{\street{Spielmannstr. 11a}, \city{Braunschweig}, \postcode{38106},  \country{Germany}}}


\abstract{Hyperdimensional (HD) computing involves encoding of baseline information into large hypervectors and repeated Boolean operations to generate the output class hypervectors which are stored in an associative memory. The classification task is then performed through similarity search operation. While prior studies have focused mostly on accelerating HD search operation using TCAMs based on emerging non-volatile memories, considering the dominant contribution of the encoder module to the energy and latency landscape specifically for complex datasets such as language recognition, DNA sequencing, etc., in this work, we propose energy- and area-efficient single FDSOI ferroelectric (Fe)FET-based logic-in-memory implementations of XOR and 3-input majority gates for N-gram HD encoders. We utilize the proposed FeFET-based encoder in a HD spam filtering accelerator and show that it outperforms the prior emerging non-volatile memory-based implementations in terms of area and energy-efficiency while exhibiting a high classification accuracy of 91.38\% on the SMS Spam Collection dataset.}


\maketitle

\section{Introduction}\label{sec1}

The frequent energy hungry data transfer involved in the traditional von-Neumann computing systems while handling computationally-intensive tasks has led to the development of alternate brain-inspired computing paradigms, such as in-memory computing, hyperdimensional (HD) computing, etc. HD computing leverages the innate property of high-dimensional vector spaces where hypervector representation of data becomes orthogonal at large dimensions  \cite{kanerva1988sparse_orthogonality_1,datta2019programmable_orthogonality_2,widdows2015reasoning_orthogonality_3}. This inherent behavior of high-dimensional hypervectors enables robust, distributed, and noise-tolerant computing. HD computing has been widely applied to several classification tasks such as DNA sequencing, language recognition, etc. \cite{imani2018hdna, rahimi2016robust_language_MATLAB}. HD computing for classification tasks consists of two modules: encoder and associative memory. The encoder module first converts the individual entries from the dataset into high-dimensional hypervectors and then applies several bit-wise boolean operations on these individual hypervectors to create hypervectors for different classes. The associative memory stores the class hypervectors generated by the encoder module and the classification is then performed by feeding the hypervector of the test dataset (again generated by the encoder module) as a query to the associative memory and searching for the nearest class hypervector (with least hamming distance). Since the HD encoder module involves frequent and large-scale Boolean operations - primarily XOR, majority, and shift on the hypervectors which are typically performed sequentially, it dominates the delay and energy landscape of the HD computing primitives specifically for complex datasets such as DNA sequencing, language recognition, etc. \cite{imani2018hdna,rahimi2016robust_language_MATLAB} and limits their throughput and energy-efficiency.

In-memory HD computing, which performs computations directly within the memory block, can minimize data movement and significantly reduce the delay \cite{FeFET_IMC_HDC_XOR_majority, PCM_XOR, HD_TCAM_sram}. While prior research has primarily focused on accelerating HD search operation in the associative memory using ternary content-addressable memories (TCAMs) \cite{HD_TCAM_sram}, the hypervector encoding stage remains a significant bottleneck for practical classification tasks involving large datasets. A fast and efficient in-memory HD computing system must also include an optimized in-memory encoding module. Moreover, by instantiating dedicated processing elements for each hypervector bit, HD computing can be parallelized across thousands of bits, achieving substantial speedups. However, this necessitates development of compact and low power Boolean logic-in-memory implementations.

Recently, emerging non-volatile memories (eNVMs) have gained significant traction for logic-in-memory implementations due to their low power consumption, high storage density, multi-bit capability, and suitability for 3D integration \cite{mulaosmanovic2021ferroelectric,PCM_review_1,PCM_review_2,RRAM_review_2,MORRAM_review_1,STTMRAM_3D,MTJ_endurance_retention}. Among these eNVMs, Ferroelectric Field-Effect Transistors (FeFETs) have emerged as the most promising candidate owing to their CMOS compatibility, fast switching speed, high endurance, and compact footprint \cite{mulaosmanovic2021ferroelectric, FeFETendurance10to11}. Furthermore, the inherent redundancy of HD computing stemming from the use of high-dimensional representations enables robust tolerance to device-level variations, noise, and partial hardware failures which makes FeFETs well-suited for HD computing architectures.

Considering the promising attributes of the FeFETs and the urgent need for compact and energy-efficient encoder module for HD computing, in this work, for the first time, we present a single (1T)-Fully Depleted Silicon-On-Insulator (FDSOI) FeFET-based logic-in-memory implementation of the XOR and majority gate which perform the dominant binding and bundling operations within the HD encoder module. The proposed designs are highly compact, occupying only 0.007 $\mu$$m^2$ per gate, and exhibit ultra-low worst-case energy of 0.41 fJ and 0.65 fJ for XOR and majority gates, respectively enabling scalable, parallel, and energy-efficient in-memory HD computing. Furthermore, to analyze the impact of the proposed encoder design at the architectural level, we develop an N-gram-based HD computing model and apply it to a real-world spam filtering task using the SMS dataset \cite{sms_spam_collection_228}, achieving a peak classification accuracy of 91.38\%. We also investigate the impact of varying the N-gram window width (N) and hypervector dimensionality (D) on the spam filtering classification accuracy for the first time.

The manuscript is organized as follows: Section~2 provides a brief introduction to HD computing, highlighting its unique attributes and the dominant operations in high-dimensional space. Different encoding schemes and search modules employed in HD computing are discussed in Section~3. Section~4 reviews the existing compute-in-memory implementations of XOR and majority gates and evaluates their feasibility and efficiency for in-memory HD computing. Section~5 describes the experimental characterization and simulation methodology including the details regarding the in-house developed compact model used in this study, along with the program and erase scheme utilized for the FDSOI FeFET. Section~6 presents the proposed single-transistor FDSOI FeFET-based XOR and majority gates and explains their working principle in detail. Section~7 outlines an $N$-gram encoding-based HD computing implementation for spam filtering based on the proposed FeFET-based encoder design. The impact of inherent variations in scaled FeFET devices on the performance of the designed in-memory logic gates is assessed in Section~8. The performance metrics of the proposed logic-in-memory implementation and the encoder modules are evaluated and benchmarked in Section~9 while conclusions are drawn in Section~10.

\section{Hyperdimensional Computing}
Hyperdimensional (HD) computing is an emerging paradigm which takes inspiration from the human brain, wherein data is represented as high-dimensional vectors—typically with tens of thousands of components \cite{FeFET_IMC_HDC_XOR_majority,HDC_review}. This high-dimensional (hypervector) representation introduces redundancy, making the system inherently robust to noise and partial failures \cite{FeFET_IMC_HDC_XOR_majority}. Furthermore, the large-dimensional hypervectors tend to be nearly orthogonal and facilitates distributed representation enabling efficient and error resilient hardware implementation \cite{kanerva1988sparse_orthogonality_1,datta2019programmable_orthogonality_2,widdows2015reasoning_orthogonality_3}.

\subsection{Encoder Module}

Hyperdimensional computing is capable of processing various forms of input data, including letters, signals, and images \cite{review_hd_ge2020classification}. For classification tasks, the encoder transforms the input data into high-dimensional vectors that are selected randomly or pseudo-randomly, with elements being independent and identically distributed (i.i.d.). These HD vectors may consist of binary, bipolar, integer, or complex-valued elements. Furthermore, the individual hypervectors are processed through a set of well-defined operations to derive the class hypervectors. The core operations in HD computing are multiplication, addition, and permutation—commonly referred to as the MAP operations.

\subsubsection{Addition/Bundling}
For a given set of input hypervectors ${I_1, I_2, \ldots, I_n}$, the point-wise addition or bundling operation produces another hypervector $Z$ such that $Z$ is maximally similar to each of the $n$ component hypervectors \cite{HDC_review}. This operation preserves the information content of all the inputs.

\begin{gather}
Z = [I_1 + I_2 + \ldots + I_n]
\end{gather}

Hypervectors with integer-valued elements typically yield higher classification accuracy albeit at the cost of increased computational complexity. To mitigate this, the resulting vector $Z$ is often thresholded and binarized to ${0,1}$ using a majority function \cite{imani2018hdna,review_hd_ge2020classification}. For binary hypervectors, a simple majority operation is sufficient to retain the essential information during the bundling process. However, implementing such large-order majority function for HD computing architectures incurs a significant area overhead and energy consumption. Therefore, there is an urgent need for an efficient logic-in-memory majority gate implementation which can significantly reduce the overall energy and footprint while simultaneously alleviating performance degradation owing to the von-Neumann bottleneck.

\subsubsection{Multiplication/Binding}
The binding operation, implemented as point-wise multiplication (XOR for binary hypervectors), encodes the association between two hypervectors by generating a new hypervector that is nearly orthogonal to both inputs.

\begin{gather}
Z = I_1 \oplus I_2
\end{gather}

For a $D$-dimensional hypervector, at least $D$ XOR gates are required to perform the binding operation. Since $D$ is generally in the order of tens of thousands, designing energy- and area-efficient XOR gates is critical for minimizing the overall energy consumption and footprint of the HD computing systems.

\subsubsection{Permutation}
Permutation ($\rho$) is a unary operation in hyperdimensional computing, typically used to mix or shuffle the elements of a hypervector. The resulting hypervector is quasi-orthogonal to the original hypervector, exhibiting a normalized Hamming distance of approximately 0.5. While permutation can be implemented by multiplying the hypervector with a permutation matrix, the hardware-friendly circular shifting is widely adopted in HD computing systems. In practice, circular shifting can be realized through intelligent mapping and routing of the encoder’s intermediate hypervector bits, enabling efficient permutation without significant energy and delay overheads \cite{FeFET_IMC_HDC_XOR_majority,kanerva2009hyperdimensional_review}.

\subsection{Encoding Schemes}
 Among different encoding strategies, the two popular schemes are record-based encoding \cite{imani2018hdna} and N-gram-based encoding \cite{imani2018hdna,rahimi2016robust_language_MATLAB}. 

\subsubsection{N-gram-based encoding}

\begin{figure}[!t]
\centering
\includegraphics[scale=0.7]{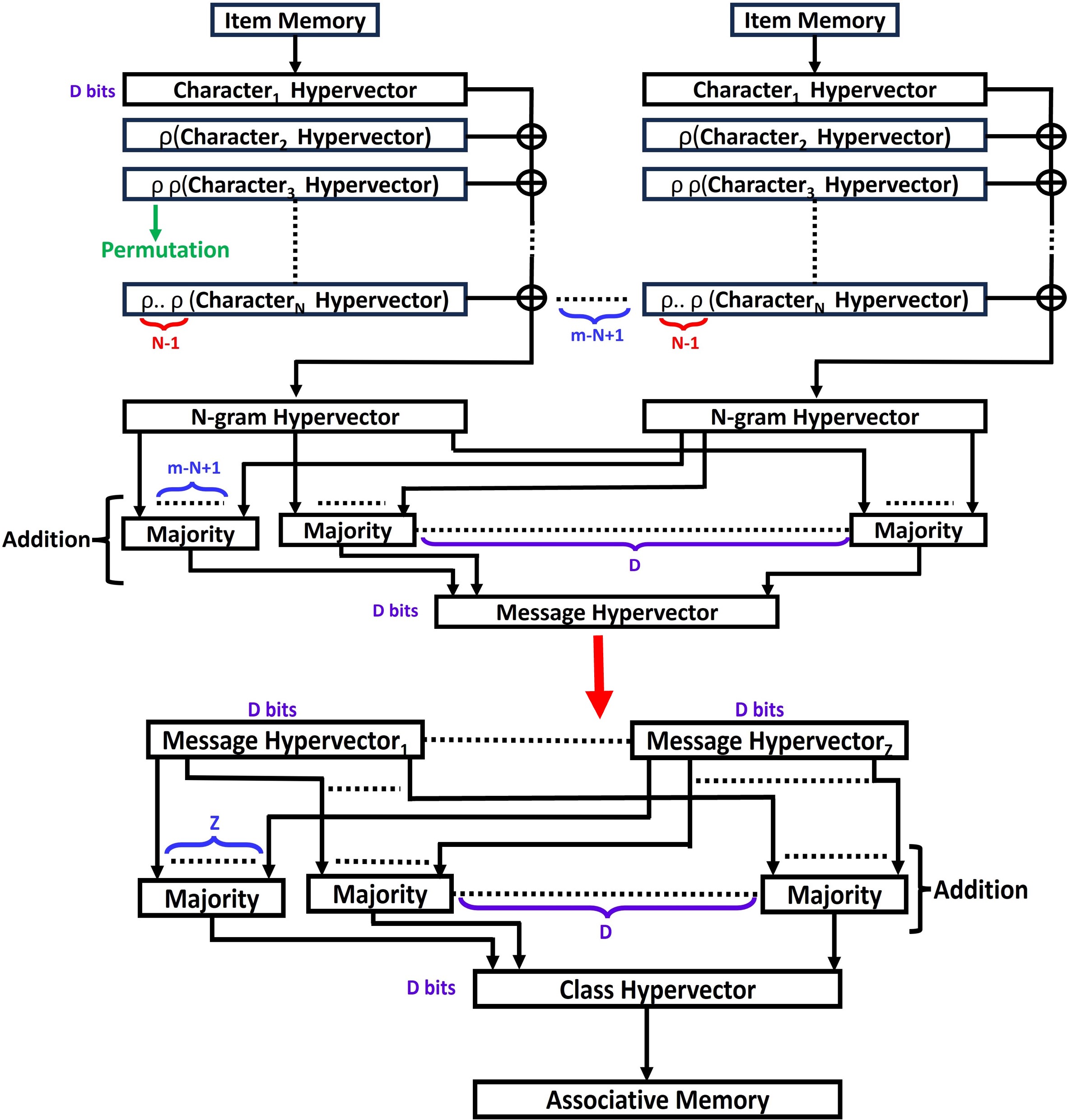}
\caption{N-gram-based encoding module.}
\label{fig1}
\end{figure}

The N-gram-based encoding module is illustrated in Fig. \ref{fig1}. In N-gram-based encoding, the input data is divided into $N$-sized windows, where $N \in {2, 3, \ldots}$. Initially, random i.i.d. hypervectors are assigned to each unique input in the dataset. For every N-gram window, the corresponding hypervectors of each input are permuted (circularly shifted and represented by $\rho$), and subsequently combined using the binding operation (XOR) to form a single hypervector that retains the characteristics of all inputs within the N-gram window.

\begin{gather}
S_1 = [L_1 \oplus \rho(L_2) \oplus \rho^2(L_3) \oplus \cdots \oplus \rho^{n-1}(L_n)]
\end{gather} Either Use L or I for representing inputs.

This process continues sequentially until all N-gram windows across the input sequence (e.g., message, signal, or image) have been processed. The resulting N-gram hypervectors ($S_1, S_2, \ldots$) are then bundled together using an element-wise majority operation to generate the final hypervector representing the entire input sequence:

\begin{gather}
S = [S_1 + S_2 + \cdots + S_{m-N+1}]
\end{gather}

where $[+]$ denotes the majority (bundling) operation and $\oplus$ represents the XOR (binding) operation. Furthermore, the hypervectors corresponding to multiple input sequences are bundled to produce the class hypervectors. This encoding scheme relies on extensive use of majority and XOR gates, highlighting the critical need for their efficient hardware realization. 

\subsubsection{Record-based encoding}

The record-based encoding scheme captures the positional information of each input in the sequence by assigning a unique identification (ID) hypervector, randomly generated and semi-orthogonal to other hypervectors, to each position in the sequence. The final hypervector corresponding to an input sequence is computed in a single step \cite{HDC_review,imani2018hdna}:

\begin{gather}
S_1 = [ID_1 \oplus L_1 + ID_2 \oplus L_2 + \cdots + ID_m \oplus L_m]
\end{gather}
\begin{gather}
\text{where } {L_1, L_2, \ldots, L_m} \in L_{\text{elements}} \nonumber
\end{gather}

where $m$ denotes the length of the longest sequence. Similar to the N-gram-based approach, the resulting sequence hypervectors are binarized via a point-wise majority operation. Although this encoding technique yields higher accuracy compared to N-gram-based encoding for classification task \cite{imani2018hdna}, it is more resource-intensive and requires the use of XOR and majority gates in the order of tens of thousands to ensure fast and efficient processing.

\subsection{Associative Memory}
\begin{figure}[!t]
\centering
\includegraphics[scale=0.14]{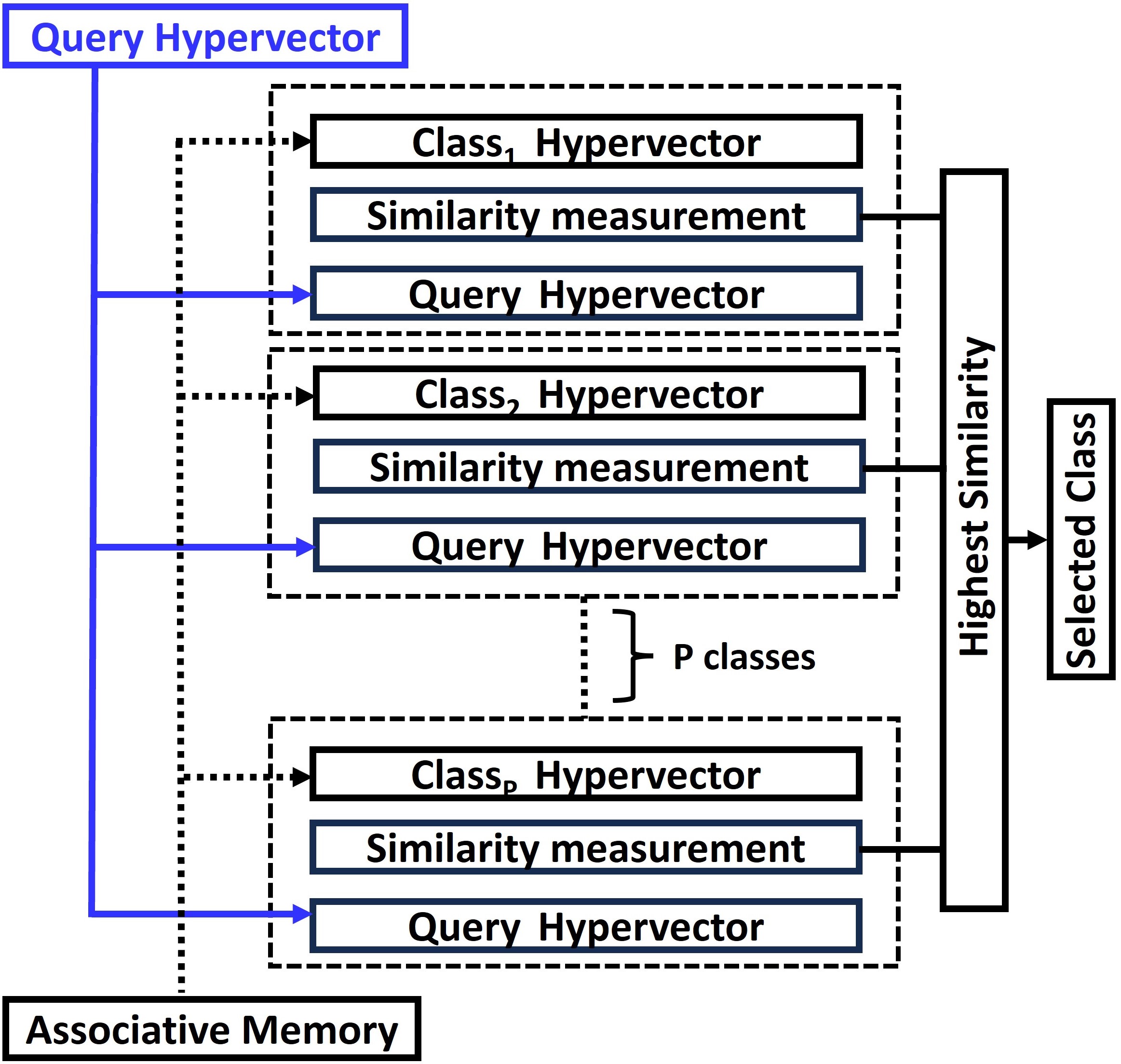}
\caption{Search module.}
\label{fig2}
\end{figure}

The class hypervectors generated by the encoder module are stored in an associative memory. During testing (inference) phase, the test data is first transformed into a new hypervector, known as the query hypervector, utilizing the same encoding scheme. Then the query hypervector is compared with each of the class hypervectors stored in the associative memory, as illustrated in Fig. \ref{fig2}. The label corresponding to the class hypervector that exhibits the maximum similarity with the query hypervector is assigned as the predicted label for the test input. To quantify the degree of similarity, either cosine similarity or Hamming distance is employed, depending on the nature of the hypervector representation (e.g., binary or real-valued).

\section{Prior Works}

Recently, several works have explored the use of emerging non-volatile memories (NVMs) for logic-in-memory implementation targeting HD computing. While single cycle logic gate implementations using voltage threshold-based memristor crossbar was proposed in \cite{FELIX}, 3D vertical ReRAM pillars were used to demonstrate multiplication-addition-permutation (MAP) operations in \cite{3D_ReRAM_MAP_IEDM}. While these RRAM-based implementations offer high area-efficiency, their operating speed and energy-efficiency are limited due to the current-based write mechanism. Furthermore, memristor arrays were also utilized to implement the majority and XOR operations in \cite{xor_memristor_searchd}. However, the energy- and time-consuming programming from the XOR array to majority array limits its practical applications.

Similarly, phase-change memories (PCMs) were also used to implement bitwise XOR operations for an N-gram-based HD encoder module in \cite{PCM_XOR}. However, the design still relied on complex digital peripheral circuits to perform hardware-intensive addition and thresholding operations. 

Recently, considering the promising attributes of FeFETs, a FeFET-based XOR gate was also proposed by exploiting the inherent AND operation between stored data and an applied drain voltage in \cite{FeFET_IMC_HDC_XOR_majority}. However, this design is complex and requires two FeFETs storing complementary inputs, with complementary voltages applied to their drain terminals, followed by a NOR gate resulting in a pseudo in-memory computing architecture with limited gains in energy and speed. Additionally, one operand for the XOR operation is stored in a D-flip-flop, which increases the energy consumption and area overhead. Also, the majority gate implementation proposed in \cite{FeFET_IMC_HDC_XOR_majority} uses a supercell consisting of a FeFET in series with a passive resistor, along with an access p-MOSFET. This configuration further increases the area overhead and requires four control signals per supercell, exacerbating routing and synchronization challenges. Moreover, the readout was performed using energy hungry two-stage differential amplifier-based buffer followed by a strongARM latch-based comparator. Furthermore, to reconcile the fast read-based XOR operation with the slower write-based majority operation, a two-step charge-based pipelined write mechanism was proposed adding to circuit complexity. Moreover, FeFET-based majority gate implementation using three independently biased gates over the channel region was also demonstrated in  \cite{husam2022compact}. However, such an architecture with multiple independent gates not only results in a limited scalability and difficult process integration but also increases routing complexity. To mitigate these issues, in this work, we propose a single FDSOI FeFET-based logic-in-memory implementation of XOR and 3-input majority gates for N-gram HD encoders.

\section{Methodology}

\begin{figure}[!t]
\centering
\includegraphics[scale=0.6]{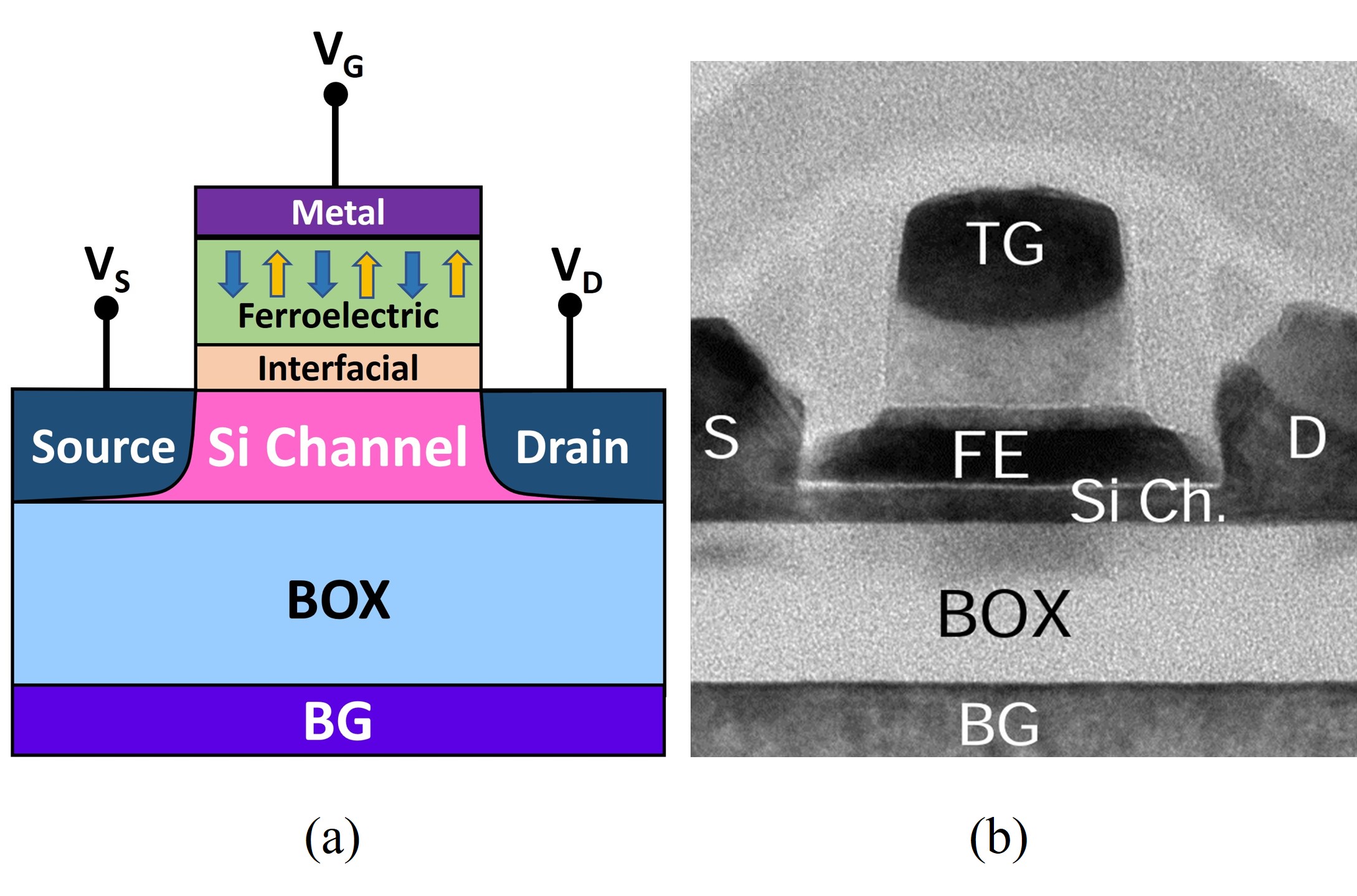}
\caption{(a) Structure of the FDSOI FeFET (b) TEM image of the FDSOI FeFET.}
\label{fig3}
\end{figure}

The FDSOI FeFETs used for proof-of-concept demonstration in our work are shown in  Fig. \ref{fig3}. These FeFETs feature a gate length of 32 nm and a gate width of 220 nm. The conventional program and erase mechanism in FeFETs involves applying positive and negative voltages at the gate terminal to switch the ferroelectric domains. However, this technique results in block erase when FeFETs are integrated in ultra-dense NAND architecture. To enable selective erase of individual FeFET devices in the array, a novel drain-erase scheme was recently proposed in \cite{wang2020drain} which involves application of positive voltage at the drain, while grounding the gate terminal \cite{wang2020drain}. Since the drain-erase scheme is most effective in devices with confined geometries such as FDSOI FeFETs \cite{wang2020drain}, erase operation was achieved by applying a 4 V drain pulse while program operation was performed using a 4 V gate pulse (gate program scheme). The details of electrical characterization methodology including program/erase and read out are as follows: To program the FeFET, a +4.5 V, 10 $\mu$s pulse was applied to the gate while source, drain, and back-gate were held at 0 V. To erase, +4.0 V, 10 $\mu$s pulses were applied simultaneously to the source and drain with the gate and back-gate at 0 V. The Write pulses were generated by a PPMU (PXIe-6570). For fast readout, two synchronized SMUs (PXIe-4143) were used: SMU1 stepped the gate voltage from -0.5 V to +2.0 V in 0.1 V increments, while SMU2 biased the drain at 0.1 V and sampled the drain current after a defined settling delay; the source and back-gate were kept at 0 V throughout. 

\begin{figure}[!t]
\centering
\includegraphics[scale=0.65]{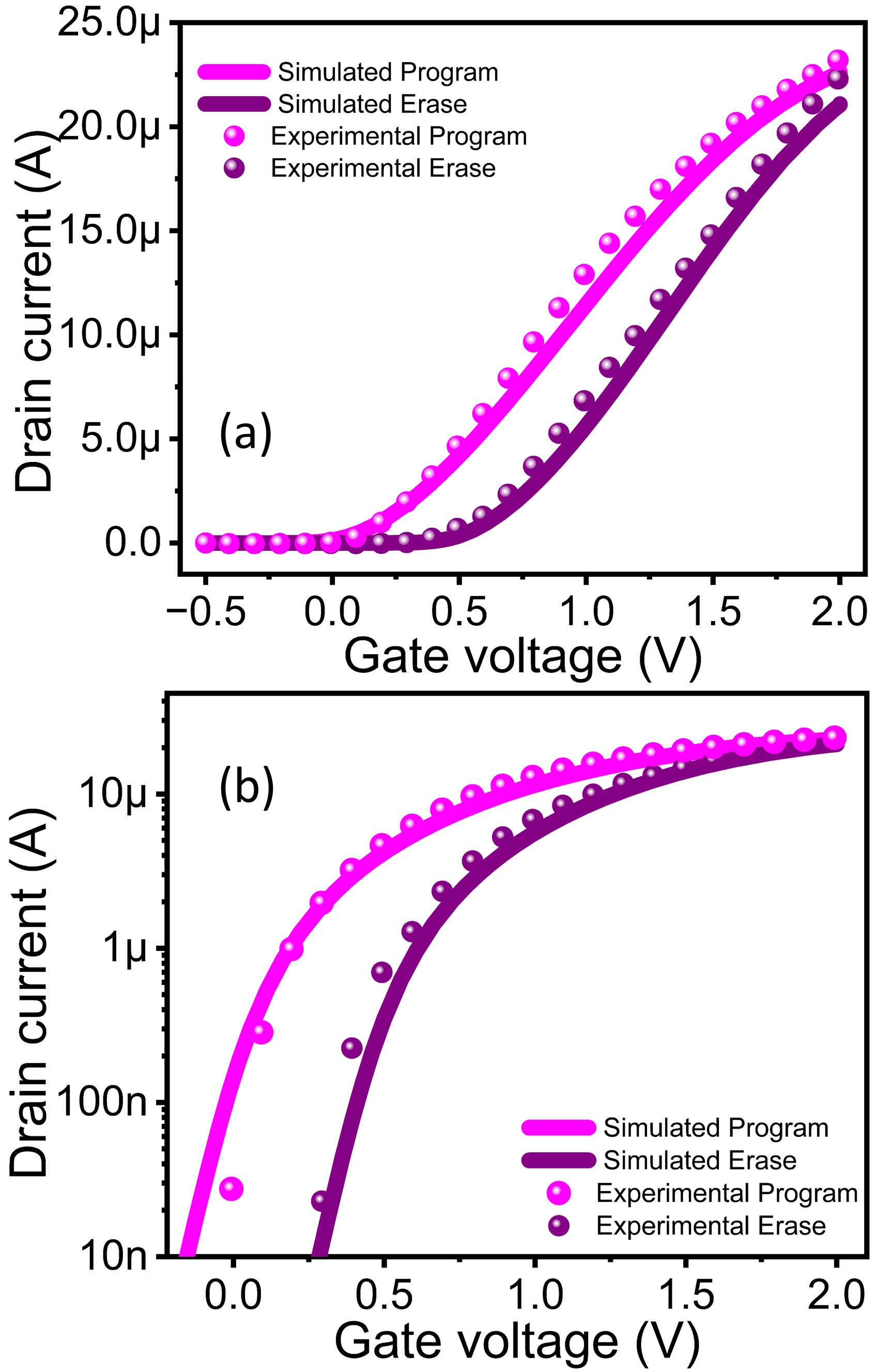}
\caption{$I_{DS}-V_{GS}$ characteristics of the FDSOI FeFET for program and erase state in (a) linear scale and (b) log scale.}
\label{fig4}
\end{figure}

To facilitate circuit and system-level estimations, we developed a SPICE-compatible compact model for the FDSOI FeFETs using the industry-standard BSIM-IMG model along with a multi-domain Preisach model for the ferroelectric layer following \cite{gaidhane2022computationally}, \cite{chatterjee2023ferroelectric}. Furthermore, the model parameters were carefully tuned to reproduce the experimentally measured program/erase characteristics of the FDSOI FeFETs. For the ferroelectric HZO layer, a saturation polarization ($P_s$) of 15 $\mu$C/cm$^2$, a remnant polarization ($P_r$) of 14.5 $\mu$C/cm$^2$, a dipole relaxation time ($\tau_v$) of 0.1 $\mu s$ and a coercive field ($E_C$) of 1 MV/cm were used. Furthermore, for the BSIM-IMG FET model, key mobility and threshold voltage parameters such as $U_0 = 11.5$ m, $DVT0 = 30$, $ETAMOB = 0.8$, and $U0MULT = 0.22$ were tuned to match the experimental data. The program and erase drain current characteristics of the developed FDSOI FeFET model shows excellent agreement with the experimental data as shown in Fig.\ref{fig4}.


\begin{figure}[!t]
\centering
\includegraphics[scale=0.2]{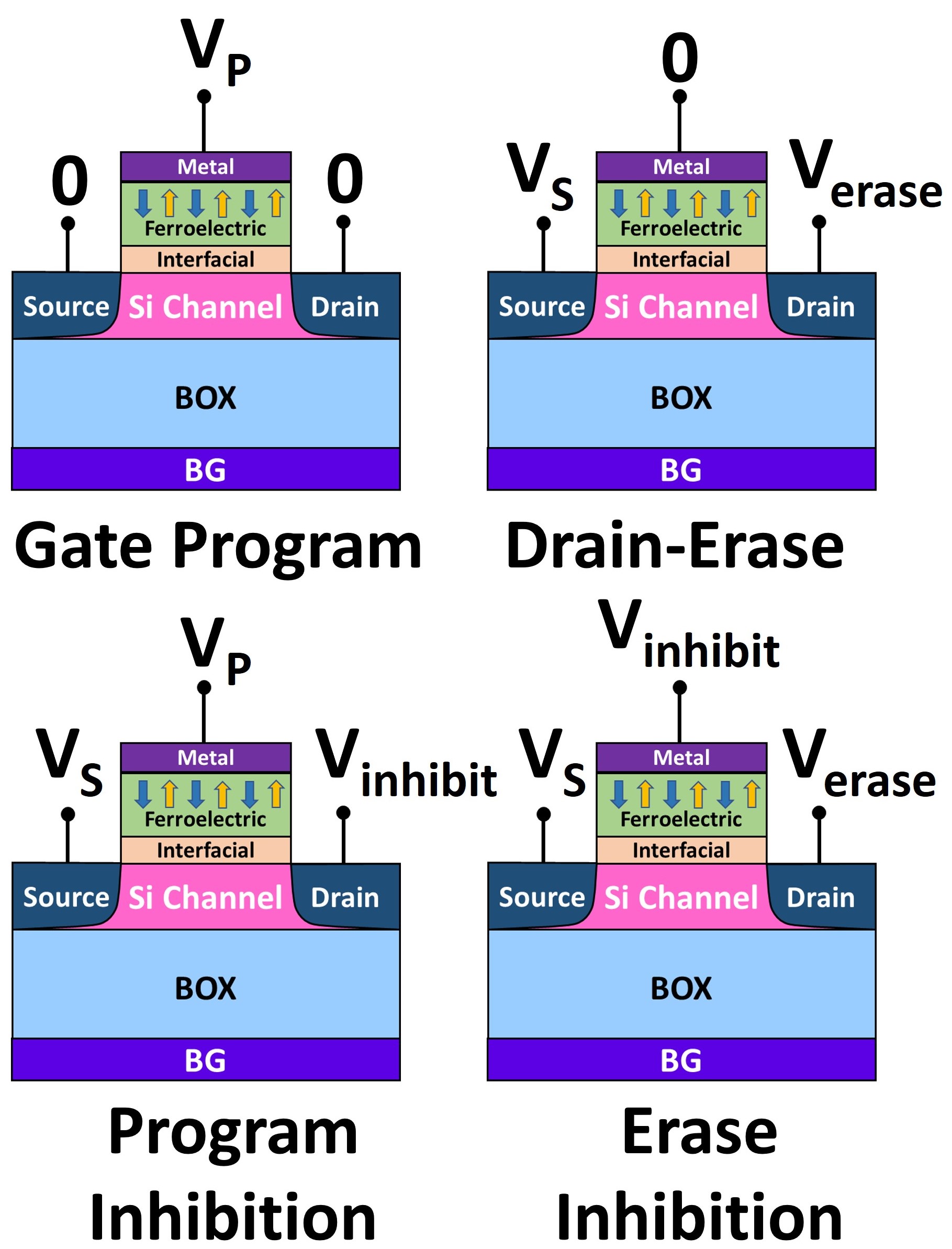}
\caption{Different pulsing conditions for program, drain-erase, program inhibition and erase inhibition in FeFET.}
\label{fig5}
\end{figure}

To design the FDSOI FeFET-based in-memory XOR and majority logic gates, we utilized four basic operations as shown in Fig. \ref{fig5}: (a) Gate Program, (b) Drain Erase, (c) Program Inhibition, and (d) Erase Inhibition. While gate program involves application of a high positive voltage ($V_{\text{Program}}$) at the gate terminal while grounding the drain and source terminals to realize a low $V_t$ state, applying a high positive voltage at the drain terminal ($V_{\text{Erase}}$) (and optionally on the source terminal ($V_S$) to improve the efficacy of the drain-erase mechanism) while grounding the gate terminal leads to a high $V_t$ state. Furthermore, application of a positive voltage to the drain and source terminals can inhibit the gate program (program inhibition) while application of a positive gate voltage can inhibit the drain erase scheme (erase inhibition), thereby preserving the inherent polarization-state of the FeFETs \cite{Arka}. In the subsequent sections, we discuss the proposed methodology for realizing single FDSOI FeFET-based logic-in-memory implementations for HD computing.

\subsection{Majority gate using single FDSOI FeFET}

\begin{figure}[!t]
\centering
\includegraphics[scale=0.3]{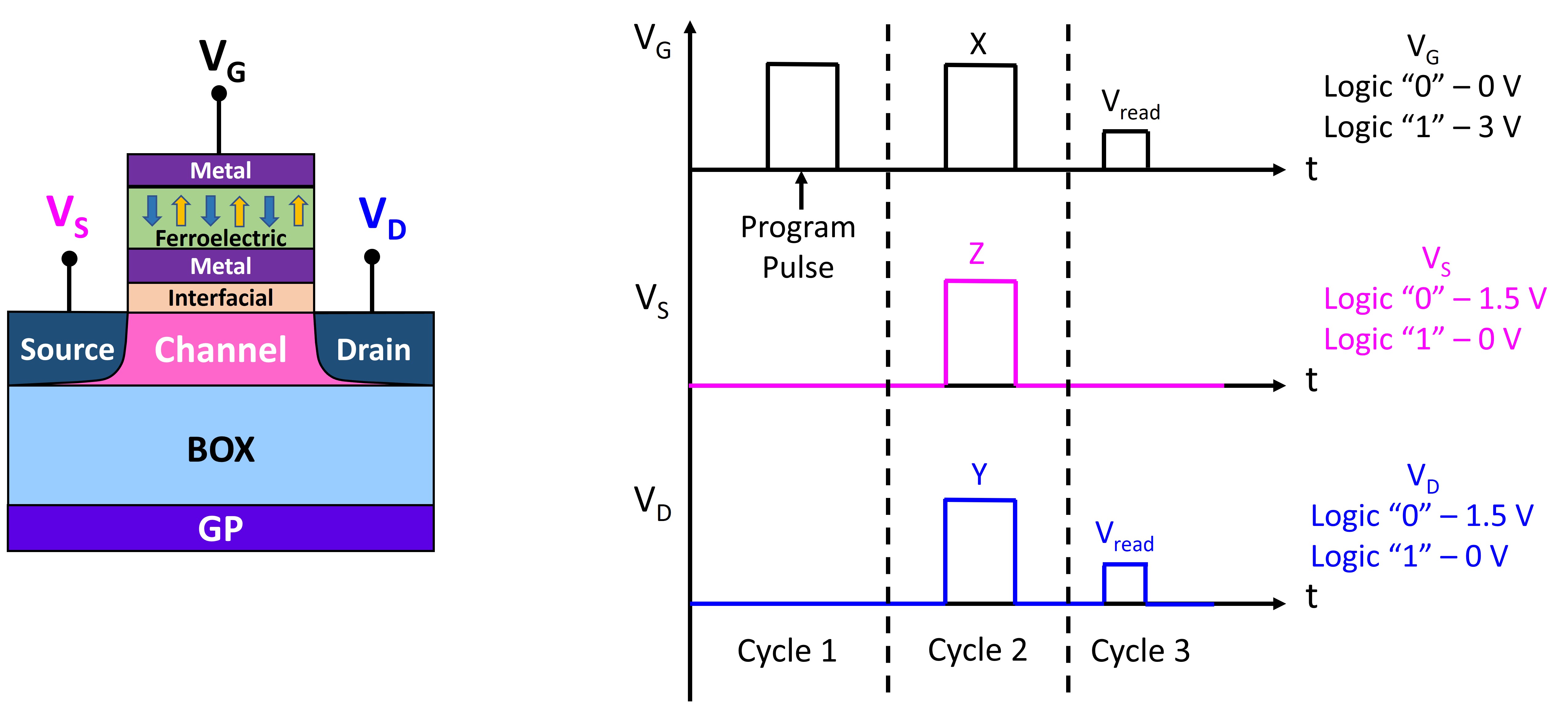}
\caption{The proposed single FDSOI FeFET-based logic-in-memory 3-input majority gate.}
\label{fig6}
\end{figure}

The majority logic gate consists of an odd number of inputs and the output is high if majority of the inputs are in logic high state, and low otherwise. We propose a single-FDSOI FeFET-based 3-input majority gate leveraging the drain-erase mechanism \cite{wang2020drain} as shown in Fig.~\ref{fig6}. The proposed majority gate uses mixed logic encoding for the inputs: while positive logic (logic ``0" = 0 V, logic ``1" = 3 V) is used for input X applied to the gate terminal, negative logic (logic ``0" = 1.5 V, logic ``1" = 0 V) is applied to the inputs Y and Z at the drain and source terminals, respectively.

\begin{figure}[!t]
\centering
\includegraphics[scale=0.9]{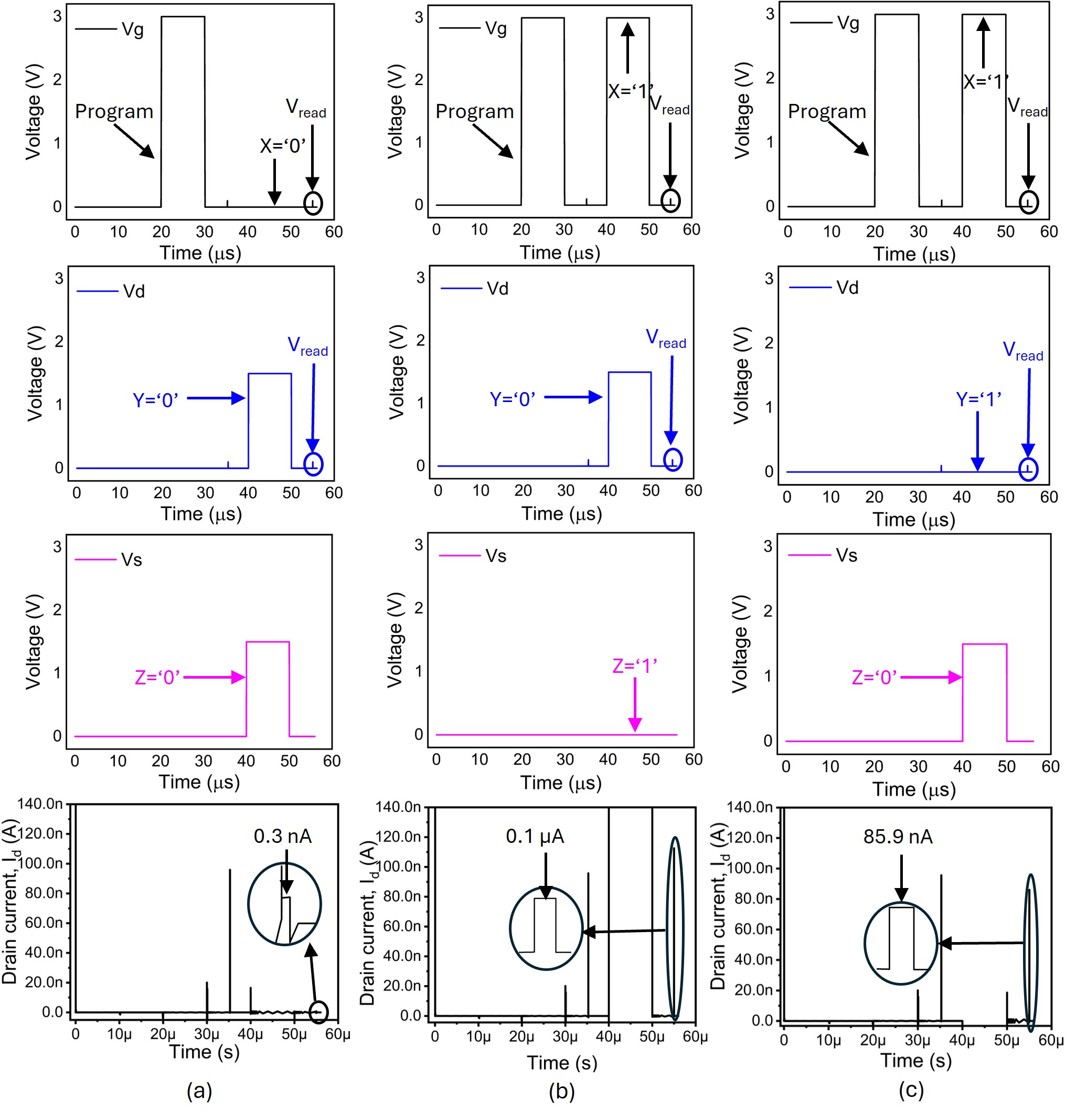}
\caption{Timing diagram for the proposed majority gate utilizing FDSOI FeFET during (a) input combination of 000 leading to drain-erase (b) input combination of 101 leading to erase-inhibition and (c) input combination of 110 leading to erase-inhibition.}
\label{fig7}
\end{figure}

The FDSOI FeFET-based majority gate implementation leverages the voltage-dependent polarization switching behavior of the ferroelectric layer, and the final output is obtained by sensing the resulting polarization state through read current measurement. First, a positive gate voltage ($V_g$) of 3 V is applied to initialize the FeFET to a low threshold voltage ($V_t$) state prior to the application of the inputs. The three inputs X, Y and Z are then applied simultaneously to the gate $V_g$, drain $V_d$ and source $V_s$ terminals. Depending on the combination of the inputs, the FeFET undergoes either full erase, partial erase, or erase inhibition. The final output is obtained by applying a low-voltage read pulse at the gate and drain terminals and measuring the read current. The truth table summarizing the output logic for all input combinations is presented in Table \ref{tab1}, and selected characteristics illustrating erase and erase-inhibition mechanisms for representative inputs (``000", ``101", ``110") are shown in Fig.~\ref{fig7}.

For input combinations such as ``001" and ``010", where only one of the $V_d$ or $V_s$ terminals is biased at 1.5 V, the channel potential is boosted somewhat inducing partial drain-erase mechanism \cite{wang2020drain} leading to a low read current corresponding to output logic ``0''. Since the drain-erase mechanism is more pronounced when both the drain ($V_d$) and source ($V_s$) terminals are biased at 1.5 V \cite{wang2020drain}, such as for input combinations ``000" and ``100", the channel potential is boosted significantly resulting in an even higher threshold voltage ($V_t$) state and a lower read current while the output logic still corresponds to ``0''. In contrast, for input patterns such as ``101" or ``110", even though $V_d$ or $V_s$ is at 1.5 V, the gate terminal is biased at 3 V effectively inhibiting the erase operation preserving the initial low $V_t$ state leading to a high read current and an output logic ``1". Similarly, for inputs ``011" and ``111", the FDSOI FeFET maintains its low $V_t$ state due to the absence of an effective erase condition, thereby retaining its initially programmed low $V_t$ corresponding to an output logic ``1".

\begin{table}
\centering
\caption{Current states for different input combinations of the proposed FDSOI FeFET-based logic-in-memory 3-input majority gate.}
\label{table6}
\setlength{\tabcolsep}{3pt}
\begin{tabular}{@{}lllllll@{}}
\toprule
X & Y & Z & Operation & $V_t$ state & Current & Output \\
\midrule
0 & 0 & 0 & Drain-erase & High $V_t$ & 0.34 nA & 0\\
0 & 0 & 1 & Partial drain-erase & High $V_t$ & 10.5 nA & 0\\
0 & 1 & 0 & Partial drain-erase & High $V_t$ & 5.62 nA & 0\\
0 & 1 & 1 & No operation & Low $V_t$ & 95.7 nA & 1\\
1 & 0 & 0 & Drain-erase & High $V_t$ & 15.8 nA & 0\\
1 & 0 & 1 & Erase inhibition & Low $V_t$ & 0.11 $\mu$A & 1\\
1 & 1 & 0 & Erase inhibition & Low $V_t$ & 85.9 nA & 1\\
1 & 1 & 1 & Program & Low $V_t$ & 0.58 $\mu$A & 1\\
\botrule
\end{tabular}
\label{tab1}
\end{table}


\subsection{XOR gate using single FDSOI FeFET}

\begin{figure}[!t]
\centering
\includegraphics[scale=0.3]{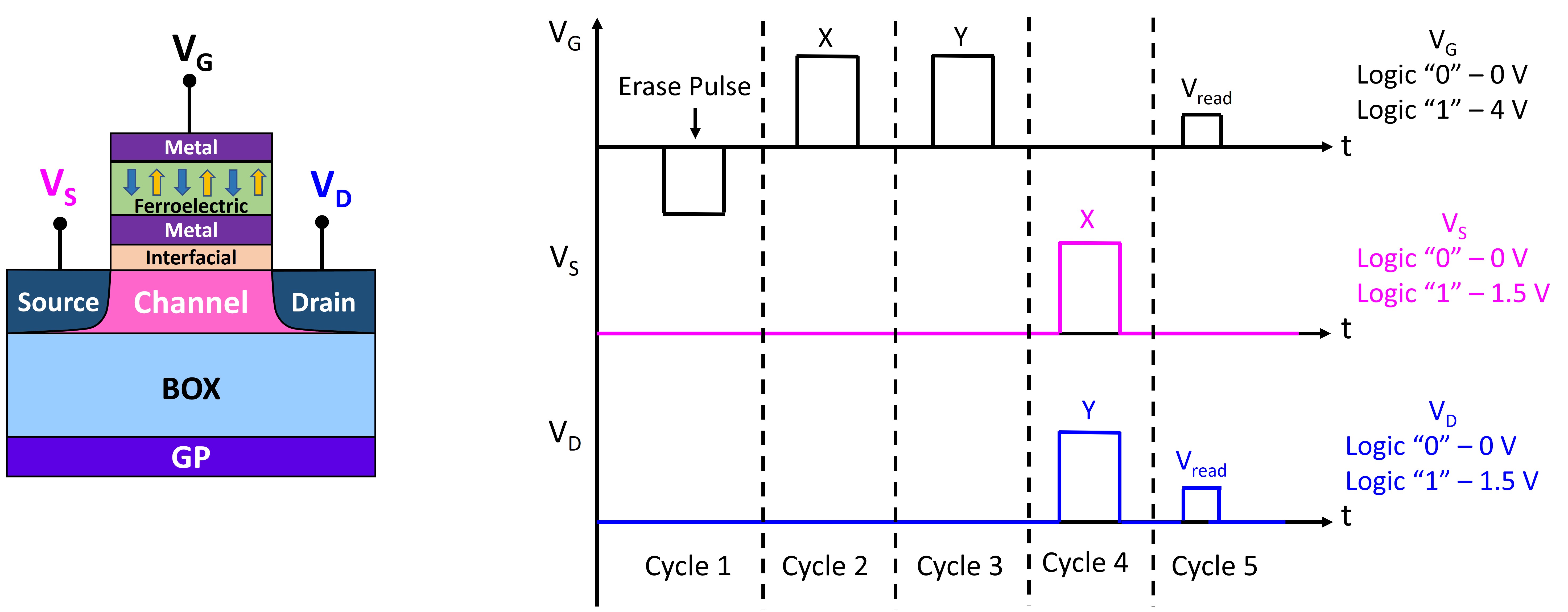}
\caption{The proposed single FDSOI FeFET-based logic-in-memory XOR gate.}
\label{fig8}
\end{figure}

\begin{table}
\centering
\caption{Current states for different input combinations of the proposed FDSOI FeFET-based logic-in-memory XOR gate.}
\label{tab2}
\setlength{\tabcolsep}{3pt}
\begin{tabular}{@{}llllll@{}}
\toprule
X & Y & Operation & $V_t$ state & Current & Output \\
\midrule
0 & 0 & No operation & High $V_t$ & 4.6 pA & 0 \\
0 & 1 & Partial drain-erase & Low $V_t$ & 0.38 $\mu$A & 1 \\
1 & 0 & Partial drain-erase & Low $V_t$ & 0.23 $\mu$A & 1 \\
1 & 1 & Drain-erase & High $V_t$ & 29.1 nA & 0 \\
\botrule
\end{tabular}
\label{tab2}
\end{table}

\begin{figure}[!t]
\centering
\includegraphics[scale=0.8]{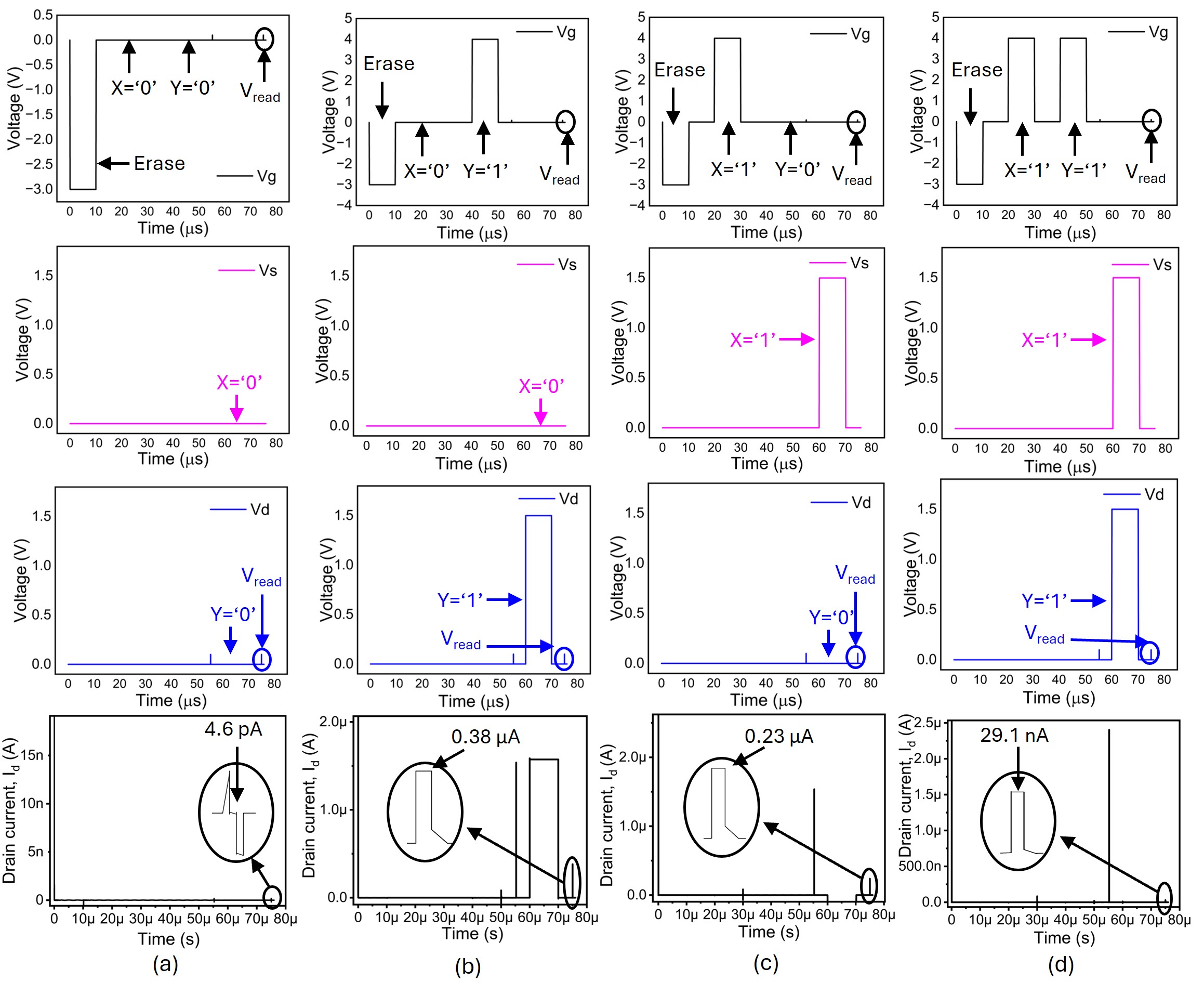}
\caption{Timing diagram for the proposed XOR gate utilizing FDSOI FeFETs for input combinations of (a) 00 (b) 01 (c) 10  and (d) 11.}
\label{fig9}
\end{figure}

  
The XOR gate produces an output logic high when the inputs are different, and a output logic low otherwise. To realize the exclusive OR operation using single FDSOI FeFET, we propose a multi-cycle approach as illustrated in Fig.\ref{fig8}. The FeFET is first initialized to a high threshold voltage ($V_t$) state by applying a negative gate voltage of -3~V. Subsequently, the two logic inputs, denoted as ``X" and ``Y", are sequentially applied to the gate terminal ($V_g$) of the FeFET. A high gate voltage of 4~V represents logic ``1", and 0~V represents logic ``0". Following the sequential application of the inputs, the inputs ``X" and ``Y" are again simultaneously applied to the source ($V_s$) and drain ($V_d$) terminals, respectively. The final output is obtained by measuring the read current. The timing diagrams and corresponding read current values for all four input combinations are shown in Fig. \ref{fig9}, with quantitative results summarized in the Table \ref{tab2}.

When both inputs are logic ``0", the FeFET remains in the high threshold voltage ($V_t$) state, corresponding to the output logic ``0". However, when both inputs are logic ``1", the FeFET gets programmed to the low $V_t$ state during cycles 2 and 3 where the inputs are applied sequentially. However, the application of high voltage to both source and drain terminals in cycle 4 significantly boosts the channel potential and fully triggers the drain-erase mechanism \cite{wang2020drain} leading to the high threshold voltage $V_t$ state. This results in a considerable reduction in the read current by three orders of magnitude leading to output logic ``0". However, when the inputs are different (i.e., either ``X" or ``Y" is logic ``1"), during the sequential application of the inputs (cycles 2 and 3), the applied gate voltage programs the FeFET into a low $V_t$ state. However, application of a high input voltage to either the source or drain terminal in cycle 4 only leads to a partial drain-erase of the FeFET leading to a moderate reduction in the read current (by one order of magnitude), thereby preserving the output logic ``1".

\begin{figure}[!t]
\centering
\includegraphics[scale=0.5]{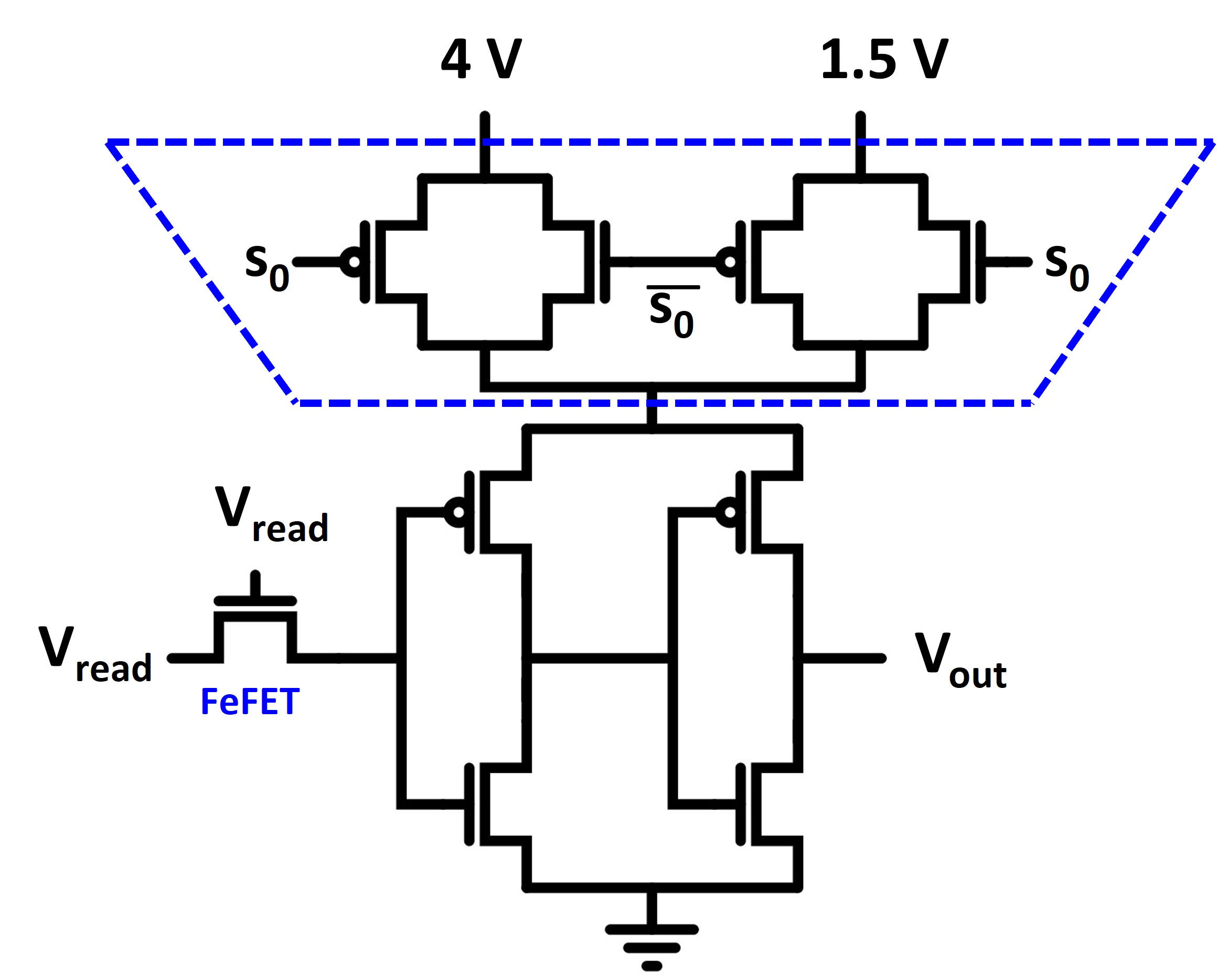}
\caption{Buffer-based read circuitry. An analog multiplexer is used to level-shift the sensed output to either 4 V or 1.5 V to ensure compatibility with subsequent stages.}
\label{fig10}
\end{figure}

For cascading the proposed single FDSOI FeFET-based logic gates to handle hypervector inputs in HD encoder, the output currents are converted into corresponding voltage levels using a compact buffer-based read circuit as shown in Fig. \ref{fig10}. The read circuitry consists of four FETs and the supply voltage is fed through shared analog multiplexer to shift the output voltage to 1.5 V or 4 V depending on the input encoding scheme for XOR and majority gate implementation.

\section{Impact of Process Variation}

Scaled FeFETs are highly susceptible to spatial (device-to-device) variations particularly in large-scale array-level implementations. To evaluate the robustness of the proposed logic-in-memory implementations based on FDSOI FeFETs in the presence of process variation, extensive Monte Carlo simulations were performed. The inherent variation in the threshold voltage ($V_t$) of the FeFETs was modeled by incorporating a Gaussian $V_t$ distribution with a 3$\sigma$ spread of 40 mV from the nominal value consistent with the experimental observations reported in \cite{swetaki_temperature_variability, Nature_FeFET_variability}.

\begin{figure}[!t]
\centering
\includegraphics[scale=0.85]{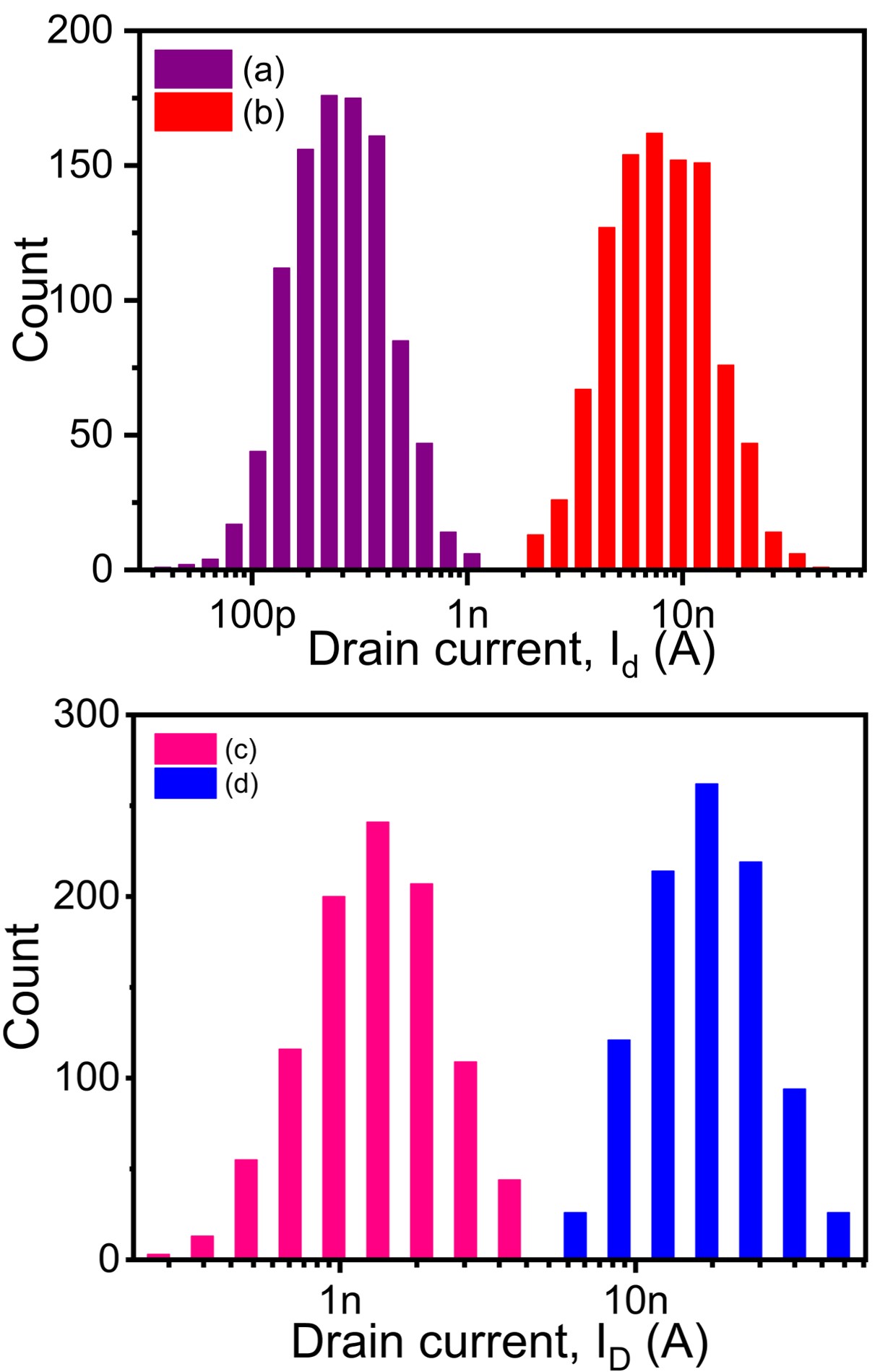}
\caption{3-sigma Monte-Carlo simulation for variation in read current with: (a) variation in $V_t$ for worst-case logic ``1" in XOR gate (b) variation in $V_t$ for worst-case logic ``0" in XOR gate (c) variation in $V_t$ for worst-case logic ``1" in majority gate (d) variation in $V_t$ for worst-case logic ``0" in majority gate.}
\label{fig11}
\end{figure}

1000 Monte Carlo simulations were performed for both the XOR and majority gate implementations based on FDSOI FeFETs. Fig.~\ref{fig11} shows the distribution of the output read current for the proposed FDSOI FeFET-based XOR and majority gates in the presence of process variation. The non-overlapping distribution and finite margin between the worst-case output logic “0” and output logic “1” clearly indicate that the proposed FDSOI FeFET-based logic-in-memory implementations are robust to the process variations which ensures reliable operation. 

\section{Spam Filtering}

For analyzing the system-level benefits of the proposed FDSOI FeFET-based logic-in-memory implementations for HD computing, we apply the proposed encoder architecture to perform spam filtering operation on the SMS Spam Collection dataset \cite{sms_spam_collection_228}. This dataset, which contains text messages labeled as either spam or ham, is first partitioned into two parts: 80\% for training and 20\% for testing. An N-gram-based encoding strategy is employed for both training and inference, as illustrated in Fig.\ref{fig1} and Fig.\ref{fig2}.

During initialization, an item memory assigns distinct, orthogonal hypervectors to each character, including letters, digits, and special symbols. This item memory acts as a lookup table, mapping each input character to a high-dimensional representation. In the training phase, the encoder module reads character sequences from the training messages, grouping them into N-gram windows. Within each window, the corresponding hypervectors are circularly shifted and then binded using the proposed FDSOI FeFET-based XOR gates to form an N-gram hypervector.

The N-gram encoding slides across the message one character at a time, generating $(m - N + 1)$ hypervectors for a message of length $m$ and window size $N$. These hypervectors are subsequently aggregated using the proposed FDSOI FeFET-based majority gates, producing a final message hypervector that preserves key statistical features of the original text. Message hypervectors corresponding to the same class (spam or ham) are further bundled to generate a class hypervector, which is stored in associative memory.

During testing, each incoming message undergoes the same N-gram-based encoding process using the proposed FDSOI FeFET-based logic gates to generate a query hypervector. This hypervector is then compared against the stored class hypervectors using Hamming distance as the similarity metric. The class with the highest similarity (lowest hamming distance) is selected, and the test message is classified accordingly as either spam or ham.

\section{Performance Metrics}

\subsection{Accuracy}

\begin{figure}[!t]
\centering
\includegraphics[scale=0.04]{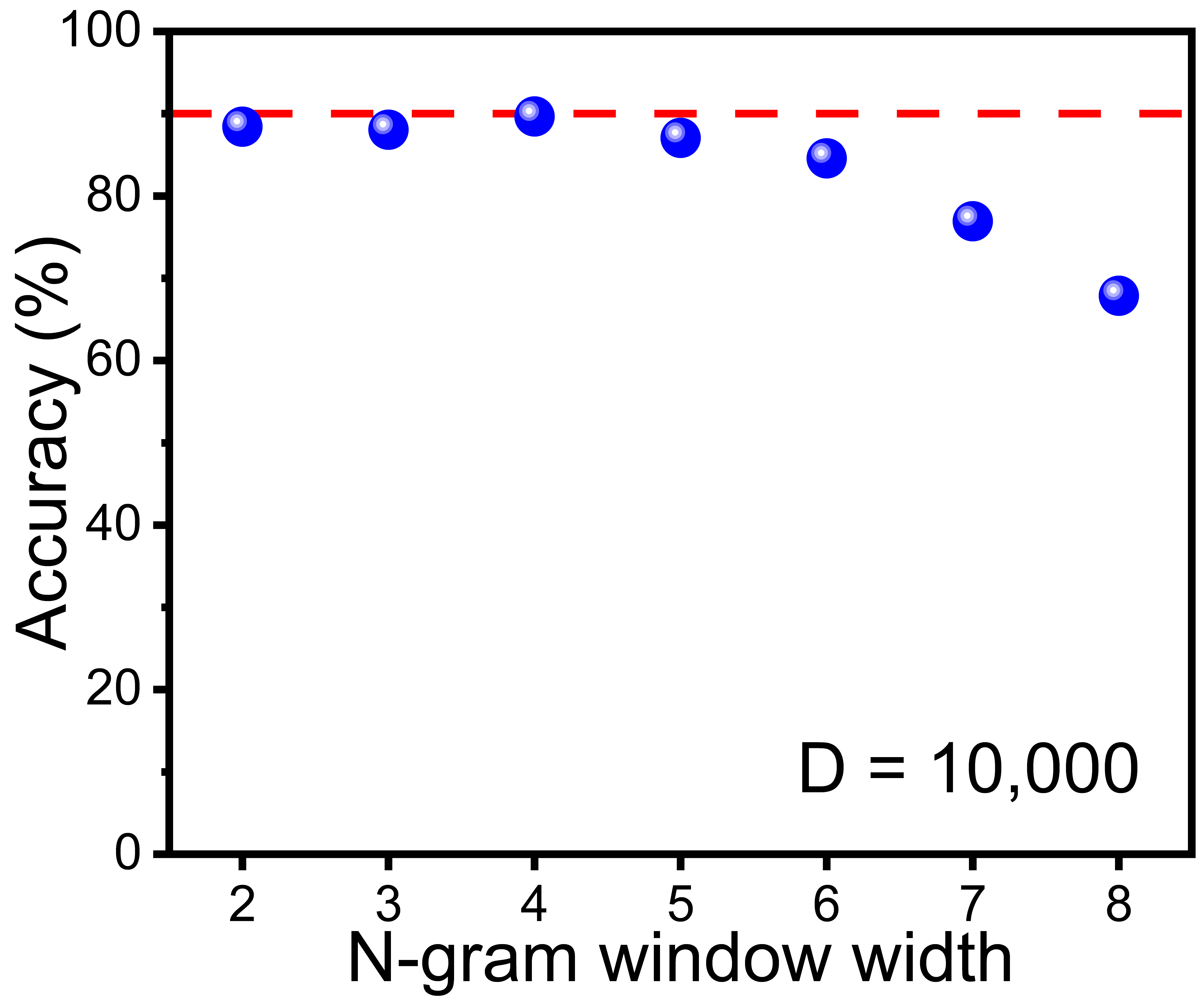}
\caption{Variation of filtering accuracy with N-gram window width (N) of hypervectors.}
\label{fig12}
\end{figure}

We performed an extensive investigation of the accuracy of the proposed HD spam filtering architecture utilizing the proposed FeFET-based logic-in-memory implementations. Since the N-gram window width (N) influences the accuracy and the encoder complexity (including the number of XOR and majority logic gates), we first investigated the impact of of N-gram window width (N) on the classification accuracy while keeping the hypervector dimensionality (D) fixed at 10,000. As shown in Fig.~\ref{fig12}, the accuracy follows a non-monotonic trend: it increases slightly with increasing N, reaching its maximum value at $N = 4$. However, increasing the N-gram window beyond $N = 4$ results in a degradation in the accuracy. This is attributed to the fact that larger N-grams capture more contextual information but they also reduce the number of training windows leading to potential overfitting and degraded generalization.

\begin{figure}[!t]
\centering
\includegraphics[scale=0.04]{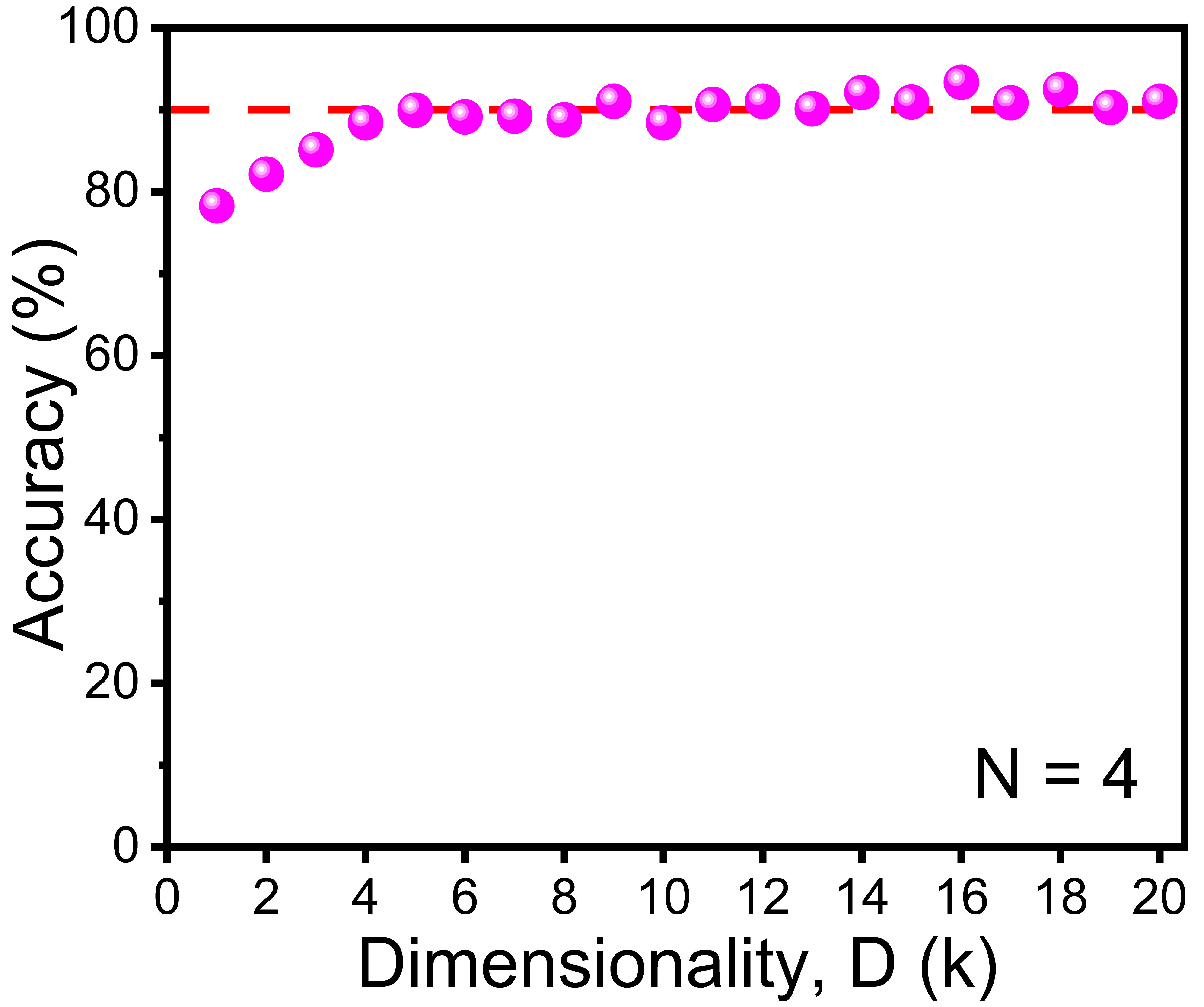}
\caption{Variation of filtering accuracy with dimensionality (D) of hypervectors.}
\label{fig13}
\end{figure}

Furthermore, we also evaluated the impact of the hypervector dimensionality D on the classification accuracy for a fixed N-gram window width of $N = 4$ (since D also influences the number of logic gates required for the encoder module). As illustrated in Fig.~\ref{fig13}, the classification accuracy improves with increasing D and saturates beyond an optimal value of $D = 9000$. As the dimensionality approaches 10000, the hypervectors become almost orthogonal, enhancing the discriminative capacity of the encoder module resulting in a higher classification accuracy \cite{kanerva1988sparse_orthogonality_1,datta2019programmable_orthogonality_2,widdows2015reasoning_orthogonality_3}. Moreover, the maximum classification accuracy observed for spam filtering using N-gram-based encoding was 91.38\% on the SMS Spam Collection dataset. The classification accuracy can be further improved by adopting record-based encoding which preserves the positional identity of elements more efficiently, however, at the cost of increased computational resource (area and energy)  \cite{Record-based_better_tha_N-gram_imani2018hierarchical}.

\subsection{Delay}

FeFETs are known to exhibit two delay components: the polarization switching delay ($t_{switch}$) and the read-after-write delay ($t_{read}$) \cite{subnsswitchingFeFETdahan2023sub, fastreadafterwritehoffmann2022fast,FeFET_switching_chargetrapping_mulaosmanovic,mulaosmanovic2021ferroelectric}. While early reports indicated a polarization switching delay of ~10 ns for FeFETs programmed using a 4 V gate voltage \cite{dunkel2017fefet} limited by the measurement setup, recent studies employing impedance-matched advanced RF probing technique have demonstrated sub-nanosecond switching, with a delay of 300 ps achieved at 4.5 V programming voltage \cite{subnsswitchingFeFETdahan2023sub}. Furthermore, even this sub-nanosecond delay is constrained by the measurement setup and the intrinsic switching time lies in the range of 1–10 ps based on the ferroelectric domain nucleation and propagation theory \cite{subnsswitchingFeFETdahan2023sub}.

Moreover, the dependence of switching time on the applied write voltage ($V_W$) has been effectively modeled using nucleation theory as follows:
\begin{gather}
t_{switch} = t_0 \cdot e^{\left(\frac{\alpha}{KT\cdot (V_W - V_0)^2}\right)}
\end{gather}
where $t_0$, $\alpha$, and $V_0$ are fitting parameters, with $t_0$ representing the minimum achievable switching time. Here, $V_0$ serves as an offset voltage, indicating that $(V_W - V_0)$ corresponds to the effective voltage drop across the ferroelectric layer. Empirical switching delays have been obtained as 60 $\mu$s, 20 ns, and 400 ps for gate voltages of 2 V, 3 V, and 4 V, respectively.

Also, accurate readout immediately after polarization switching is hindered by parasitic charge trapping effects \cite{chargetrapping1yurchuk2016charge,chargetrapping2deng2020examination,chargetrapping3wang2021standby,chargetrapping4kuk2021comprehensive}. Therefore, a minimum read-after-write delay of ~10 ns is typically required to ensure reliable sensing of the polarization-state of FeFETs \cite{fastreadafterwritehoffmann2022fast}.

Based on the switching delay and the read-after-write delays for the FeFETs, the total delay for the proposed FDSOI FeFET-based majority logic implementation is estimated as 50 ns (sum of two write operations at gate voltage of 3 V and the read-after-write delay). Similarly, the total delay for the proposed FDSOI FeFET-based XOR gate is calculated as 31.2 ns for the XOR gate (four write operations—one at gate voltage of 3 V, three at gate voltage of 4 V and the read-after-write delay). Further improvements in the write speed can be achieved by scaling down the FeFET and reducing the interfacial layer thickness relative to the ferroelectric layer, thereby enabling ultra-fast polarization switching \cite{subnsswitchingFeFETdahan2023sub}.

\subsection{Energy}

We have also analyzed the energy consumption for the proposed FDSOI FeFET-based logic-in-memory implementations for all possible input combinations to identify the worst-case energy consumption. The worst-case energy consumption is 0.41 fJ and 0.65 fJ for the proposed FDSOI FeFET-based XOR and majority gate implementations, respectively. Furthermore, the worst-case energy consumed by the HD encoder module is estimated by calculating the total number of XOR and majority gates required and multiplying them with their respective worst-case energy consumptions (which is a rare event). Considering an average email length of $m=60$ characters (obtained from the SMS spam filtering training dataset), an $N$-gram window width of $N=4$, total number of messages in training set under class spam and ham $Z=2000$, and hypervector dimensionality $D=10{,}000$, the total number of XOR gates is
\[
D \times (m-N+1) = 10{,}000 \times (60-4+1) = 5.7 \times 10^{5}.
\]
Similarly, the total number of 3-input majority gates is
\[
D \times (m-N+1) + D \times Z = 10{,}000 \times (60-4+1) + 10{,}000 \times 2,000 = 2 \times 10^{7}.
\]
Therefore, the worst-case energy consumed by the encoder module is
\[
(5.7 \times 10^{5} \times 0.41 \, \text{fJ}) + (2 \times 10^{7} \times 0.65 \, \text{fJ}) = 13.23 \, \text{nJ}.
\]

\subsection{Area}

Owing to the use of single FDSOI FeFET, the proposed logic-in-memory implementations for the HD encoder module exhibit exceptional area efficiency. Furthermore, the total footprint of the HD encoder module is obtained by multiplying the total number of XOR and 3-input majority gates with the footprint per gate. 

We have also benchmarked the proposed FDSOI FeFET-based logic-in-memory implementations against other state-of-the-art logic-in-memory realizations of XOR and majority gates in Table \ref{tab3}. Table 1 clearly indicates that the proposed FDSOI FeFET-based implementations achieve superior performance in terms of energy consumption and footprint. Although the FeFET-based logic-in-memory design in \cite{FeFET_IMC_HDC_XOR_majority} reported a faster operation, the inherent switching delay and read-after-write delays associated with FeFETs were not considered. Additionally, the implementation in \cite{FeFET_IMC_HDC_XOR_majority} employs a series resistor with the FeFET to address the ON-current variability. However, integrating such large passive resistors—particularly using polysilicon—incurs substantial area overhead, thereby compromising the overall area-efficiency.

\begin{table}
\centering
\caption{Performance Benchmarking}
\label{tab3}
\setlength{\tabcolsep}{3pt}
\begin{tabular}{p{55pt}p{40pt}p{40pt}p{40pt}p{40pt}p{50pt}p{52pt}p{52pt}}
\toprule
Implementation &
XOR delay (ns) & 
XOR energy (fJ) & 
3-input majority delay (ns) & 
3-input majority energy (fJ) & 
Area ($\mu m^2$/ bit) & 
Spam filtering HD encoder energy (nJ) &
Spam filtering HD encoder area ($mm^2$) \\
\midrule
Memristor-based \cite{FELIX} & $\sim$3 & 34.97 & $\sim$3 & 65.65 & - & 78.59 & - \\
PCM-based \cite{PCM_XOR} & 2.8 \cite{FeFET_IMC_HDC_XOR_majority} & 9.8 \cite{FeFET_IMC_HDC_XOR_majority} & - & - & 207 & - & 358.11 \\
FeFET-based \cite{FeFET_IMC_HDC_XOR_majority} &  0.557 \textsuperscript{*} & 6.21 & 1.7 \textsuperscript{*} & 24.9 & \dag & 21.77 & - \\
FeFET-based \cite{Musaib_majority} & - & - & - & 1.9$\times10^3$ & -  & - & - \\
Flash-based \cite{swaroopflashcim}  & 21.07 & 30.16 & - & - & 46.28 & - & 80.06 \\
This work & 31.2 & 0.41 & 50 & 0.65 & 0.007 & 13.23 & 0.014 \\
\botrule
\end{tabular}
\label{tab3}
\begin{center}
{\footnotesize * The write delay and the read after write delay associated with FeFETs are not considered.} \\
{\footnotesize \dag The designed majority gate incorporates a series resistor, which occupies a substantial area when implemented in polysilicon technology.} \\
\end{center}
\end{table}

\subsection{Endurance}

Prior studies \cite{100nshighpressureannealingnguyen2021wakeup,highendurance2tan2021ferroelectric,highendurance4de2021ultra,highendurance3sharma2020high,high_endurance_nature_review_10_10} have shown that FeFETs can achieve write endurance exceeding $10^{10}$ cycles by employing techniques such as high-pressure annealing, the integration of high-permittivity SiN\textsubscript{x} interfacial layers, channel-last process flows, and HF/$H_2O_2$ surface treatments during fabrication. In the worst-case scenario, the proposed FDSOI FeFET-based XOR gate involves four switching events while the majority gate implementation only requires two switching events. Therefore, a single FDSOI FeFET is expected to sustain at least $10^{10}/4 = 2.5 \times 10^{9}$ computation cycles for the XOR gate implementation and $10^{10}/2 = 5 \times 10^{9}$ cycles for the majority gate implementation.

\section{Conclusion}

In this work, we present single FDSOI FeFET-based logic-in-memory implementations of XOR and majority logic gates for encoder module of HD computing accelerators which otherwise dominate the area and energy landscape. We applied the proposed logic-in-memory implementations to an N-gram-based HD computing architecture for spam filtering on SMS dataset and showed that it outperforms the prior emerging non-volatile memory-based implementations in terms of area and energy. Our results may provide the incentive for experimental realization of in-memory HD computing accelerators.

\backmatter


\bibliography{sn-bibliography}

@article{wang2020drain,
  title={Drain--erase scheme in ferroelectric field-effect transistor—Part I: Device characterization},
  author={Wang, P. and Wang, Z. and Shim, W. and Hur, J. and Datta, S. and Khan, A. I. and Yu, S.},
  journal={IEEE Transactions on Electron Devices},
  volume={67},
  number={3},
  pages={pp. 955--961},
  month={2},
  year={2020},
  publisher={IEEE},
  doi={10.1109/TED.2020.2969401}
}

@article{gaidhane2022computationally,
  title={A computationally efficient compact model for ferroelectric switching with asymmetric non-periodic input signals},
  author={Gaidhane, A. D. and Dangi, R. and Sahay, S. and Verma, A. and Chauhan, Y. S.},
  journal={IEEE Transactions on Computer-Aided Design of Integrated Circuits and Systems},
  volume={42},
  number={5},
  pages={pp. 1634--1642},
  month={9},
  year={2022},
  publisher={IEEE},
  doi={10.1109/TCAD.2022.3203956}
}

@article{chatterjee2023ferroelectric,
  title={Ferroelectric FDSOI FET modeling for memory and logic applications},
  author={Chatterjee, S. and Kumar, S. and Gaidhane, A. and Dabhi, C. K. and Chauhan, Y. S. and Amrouch, H.},
  journal={Solid-State Electronics},
  volume={200},
  pages={pp. 108554},
  month={2},
  year={2023},
  publisher={Elsevier},
  doi={10.1016/j.sse.2022.108554}
}

@article{high_endurance_nature_review_10_10,
  title={The future of ferroelectric field-effect transistor technology},
  author={Khan, Asif Islam and Keshavarzi, Ali and Datta, Suman},
  journal={Nature Electronics},
  volume={3},
  number={10},
  pages={588--597},
  year={2020},
  publisher={Nature Publishing Group},
  doi={10.1038/s41928-020-00492-7}
}

@article{chargetrapping1yurchuk2016charge,
  title={Charge-trapping phenomena in HfO 2-based FeFET-type nonvolatile memories},
  author={Yurchuk, E. and M{\"u}ller, J. and M{\"u}ller, S. and Paul, J. and Pe{\v{s}}i{\'c}, M. and van Bentum, R. and Schroeder, U. and Mikolajick, T.},
  journal={IEEE Transactions on Electron Devices},
  volume={63},
  number={9},
  pages={pp. 3501--3507},
  month={7},
  year={2016},
  publisher={IEEE},
  doi={10.1109/TED.2016.2588439}}

@INPROCEEDINGS{chargetrapping2deng2020examination,
  author={Deng, Shan and Jiang, Zhouhang and Dutta, Sourav and Ye, Huacheng and Chakraborty, Wriddhi and Kurinec, Santosh and Datta, Suman and Ni, Kai},
  booktitle={2020 IEEE International Electron Devices Meeting (IEDM)}, 
  title={Examination of the Interplay Between Polarization Switching and Charge Trapping in Ferroelectric FET}, 
  year={2020},
  volume={},
  number={},
  pages={4.4.1-4.4.4},
  doi={10.1109/IEDM13553.2020.9371999}}

@inproceedings{chargetrapping3wang2021standby,
  title={Standby bias improvement of read after write delay in ferroelectric field effect transistors},
  author={Wang, Z. and Tasneem, N. and Hur, J. and Chen, H. and Yu, S. and Chern, W. and Khan, A.},
  booktitle={2021 IEEE International Electron Devices Meeting (IEDM)},
  pages={pp. 19--3},
  month={12},
  year={2021},
  organization={IEEE},
  doi={10.1109/IEDM19574.2021.9720502}
}

@inproceedings{chargetrapping4kuk2021comprehensive,
  title={Comprehensive understanding of the HZO-based n/pFeFET operation and device performance enhancement strategy},
  author={Kuk, S. H. and Han, S. M. and Kim, B. H. and Baek, S. H. and Han, J. H. and Kim, S. h.},
  booktitle={2021 IEEE International Electron Devices Meeting (IEDM)},
  pages={pp. 33.6.1--33.6.4},
  month={12},
  year={2021},
  organization={IEEE},
  doi={10.1109/IEDM19574.2021.9720642}
}

@article{fastreadafterwritehoffmann2022fast,
  title={Fast read-after-write and depolarization fields in high endurance n-type ferroelectric FETs},
  author={Hoffmann, M. and Tan, A. J. and Shanker, N. and Liao, Yu-Hung and Wang, Li-Chen and Bae, Jong-Ho and Hu, C. and Salahuddin, S.},
  journal={IEEE Electron Device Letters},
  volume={43},
  number={5},
  pages={pp. 717--720},
  month={3},
  year={2022},
  publisher={IEEE},
  doi={10.1109/LED.2022.3163354}
}

@article{100nshighpressureannealingnguyen2021wakeup,
  title={Wakeup-free and endurance-robust ferroelectric field-effect transistor memory using high pressure annealing},
  author={Nguyen, M. C. and Kim, S. and Lee, K. and Yim, J. Y. and Choi, R. and Kwon, D.},
  journal={IEEE Electron Device Letters},
  volume={42},
  number={9},
  pages={pp. 1295--1298},
  month={7},
  year={2021},
  publisher={IEEE},
  doi={10.1109/LED.2021.3096248}
}

@article{highendurance2tan2021ferroelectric,
  title={Ferroelectric HfO 2 memory transistors with high-$\kappa$ interfacial layer and write endurance exceeding 10 10 cycles},
  author={Tan, A. J. and Liao, Y. H. and Wang, L. C. and Shanker, N. and Bae, J. H. and Hu, C. and Salahuddin, S.},
  journal={IEEE Electron Device Letters},
  volume={42},
  number={7},
  pages={pp. 994--997},
  month={5},
  year={2021},
  publisher={IEEE},
  doi={10.1109/LED.2021.3083219}
}

@inproceedings{highendurance3sharma2020high,
  title={High speed memory operation in channel-last, back-gated ferroelectric transistors},
  author={Sharma, A. A. and Doyle, B. and Yoo, H. J. and Tung, I-Cheng and Kavalieros, J. and Metz, M. V. and Reshotko, M. and Majhi, P. and B. H., T. and Chen, Y. J.},
  booktitle={2020 IEEE International Electron Devices Meeting (IEDM)},
  pages={pp. 18--5},
  month={12},
  year={2020},
  organization={IEEE},
  doi={10.1109/IEDM13553.2020.9371940}
}

@inproceedings{highendurance4de2021ultra,
  title={Ultra-low power robust 3bit/cell Hf 0.5 Zr 0.5 O 2 ferroelectric FinFET with high endurance for advanced computing-in-memory technology},
  author={De, S. and Lu, D. D. and Le, H. H. and Mazumder, S. and Lee, Y. J. and Tseng, W. C. and Qiu, B. H. and Baig, M. A. and Sung, P. J. and Su, C. J. },
  booktitle={2021 symposium on VLSI technology},
  pages={pp. 1--2},
  month={6},
  year={2021},
  organization={IEEE},
}

@article{subnsswitchingFeFETdahan2023sub,
  title={Sub-Nanosecond Switching of Si: HfO2 Ferroelectric Field-Effect Transistor},
  author={Dahan, M. M. and Mulaosmanovic, H. and Levit, O. and D\"unkel, S. and Beyer, S. and Yalon, E.},
  journal={Nano Letters},
  volume={23},
  number={4},
  pages={pp. 1395--1400},
  month={2},
  year={2023},
  publisher={ACS Publications},
  doi={10.1021/acs.nanolett.2c04706}
}

@book{kanerva1988sparse_orthogonality_1,
  title={Sparse distributed memory},
  author={Kanerva, Pentti},
  year={1988},
  publisher={MIT press},
  address   = {Cambridge, MA}
}

@article{datta2019programmable_orthogonality_2,
  title={A programmable hyper-dimensional processor architecture for human-centric IoT},
  author={Datta, Sohum and Antonio, Ryan AG and Ison, Aldrin RS and Rabaey, Jan M},
  journal={IEEE Journal on Emerging and Selected Topics in Circuits and Systems},
  volume={9},
  number={3},
  pages={439--452},
  year={2019},
  publisher={IEEE},
  doi={10.1109/JETCAS.2019.2935464}
}

@article{widdows2015reasoning_orthogonality_3,
  title={Reasoning with vectors: A continuous model for fast robust inference},
  author={Widdows, Dominic and Cohen, Trevor},
  journal={Logic Journal of the IGPL},
  volume={23},
  number={2},
  pages={141--173},
  year={2015},
  publisher={Oxford University Press},
  doi={10.1093/jigpal/jzu028}
}

@inproceedings{imani2018hdna,
  title={Hdna: Energy-efficient dna sequencing using hyperdimensional computing},
  author={Imani, Mohsen and Nassar, Tarek and Rahimi, Abbas and Rosing, Tajana},
  booktitle={2018 IEEE EMBS International Conference on Biomedical \& Health Informatics (BHI)},
  pages={271--274},
  year={2018},
  organization={IEEE},
  doi={10.1109/BHI.2018.8333421}
}

@article{review_hd_ge2020classification,
  title={Classification using hyperdimensional computing: A review},
  author={Ge, Lulu and Parhi, Keshab K},
  journal={IEEE Circuits and Systems Magazine},
  volume={20},
  number={2},
  pages={30--47},
  year={2020},
  publisher={IEEE},
  doi={10.1109/MCAS.2020.2988388}
}

@inproceedings{rahimi2016robust_language_MATLAB,
  title={A robust and energy-efficient classifier using brain-inspired hyperdimensional computing},
  author={Rahimi, Abbas and Kanerva, Pentti and Rabaey, Jan M},
  booktitle={Proceedings of the 2016 international symposium on low power electronics and design},
  pages={64--69},
  year={2016},
  doi={10.1145/2934583.2934624}
}

@inproceedings{husam2022compact,
  title={Compact ferroelectric programmable majority gate for compute-in-memory applications},
  author={Deng, Shan and Benkhelifa, Mahdi and Thomann, Simon and Faris, Zubair and Zhao, Zijian and Huang, Tzu-Jung and Xu, Yixin and Narayanan, Vijaykrishnan and Ni, Kai and Amrouch, Hussam},
  booktitle={2022 International Electron Devices Meeting (IEDM)},
  pages={36--7},
  year={2022},
  organization={IEEE},
  doi={10.1109/IEDM45625.2022.10019400}
}

@misc{sms_spam_collection_228,
  author       = {Almeida,Tiago and Hidalgo,Jos},
  title        = {{SMS Spam Collection}},
  year         = {2012},
  howpublished = {UCI Machine Learning Repository},
  doi={10.24432/C5CC84}
}

@ARTICLE{Arka,
  author={Chakraborty, Arka and Rafiq, Musaib and Zarkob, Yawar Hayat and Chauhan, Yogesh Singh and Sahay, Shubham},
  journal={IEEE Transactions on Circuits and Systems I: Regular Papers}, 
  title={Ferroelectric FET-Based Bayesian Inference Engine for Disease Diagnosis}, 
  year={2025},
  volume={},
  number={},
  pages={1-13},
  doi={10.1109/TCSI.2025.3533044}}

@article{HD_TCAM_sram,
  title={Hw/sw co-design for reliable tcam-based in-memory brain-inspired hyperdimensional computing},
  author={Thomann, Simon and Genssler, Paul R and Amrouch, Hussam},
  journal={IEEE Transactions on Computers},
  volume={72},
  number={8},
  pages={2404--2417},
  year={2023},
  publisher={IEEE},
  doi={10.1109/TC.2023.3248286}
}

@article{swaroopflashcim,
  title={Satisfiability Attack-Resilient Camouflaged Multiple Multivariable Logic-in-Memory Exploiting 3D NAND Flash Array},
  author={Swaroop, Bhogi Satya and Saxena, Ayush and Sahay, Shubham},
  journal={IEEE Transactions on Circuits and Systems I: Regular Papers},
  year={2023},
  publisher={IEEE},
  doi={0.1109/TCSI.2023.3326332}
}

@INPROCEEDINGS{FELIX,
  author={Gupta, Saransh and Imani, Mohsen and Rosing, Tajana},
  booktitle={2018 IEEE/ACM International Conference on Computer-Aided Design (ICCAD)}, 
  title={FELIX: Fast and Energy-Efficient Logic in Memory}, 
  year={2018},
  volume={},
  number={},
  pages={1-7},
  doi={10.1145/3240765.3240811}}

@INPROCEEDINGS{3D_ReRAM_MAP_IEDM,
  author={Li, Haitong and Wu, Tony F. and Rahimi, Abbas and Li, Kai-Shin and Rusch, Miles and Lin, Chang-Hsien and Hsu, Juo-Luen and Sabry, Mohamed M. and Eryilmaz, S. Burc and Sohn, Joon and Chiu, Wen-Cheng and Chen, Min-Cheng and Wu, Tsung-Ta and Shieh, Jia-Min and Yeh, Wen-Kuan and Rabaey, Jan M. and Mitra, Subhasish and Wong, H.-S. Philip},
  booktitle={2016 IEEE International Electron Devices Meeting (IEDM)}, 
  title={Hyperdimensional computing with 3D VRRAM in-memory kernels: Device-architecture co-design for energy-efficient, error-resilient language recognition}, 
  year={2016},
  volume={},
  number={},
  pages={16.1.1-16.1.4},
  doi={10.1109/IEDM.2016.7838428}}

@article{PCM_XOR,
  title={In-memory hyperdimensional computing},
  author={Karunaratne, Geethan and Le Gallo, Manuel and Cherubini, Giovanni and Benini, Luca and Rahimi, Abbas and Sebastian, Abu},
  journal={Nature Electronics},
  volume={3},
  number={6},
  pages={327--337},
  year={2020},
  publisher={Nature Publishing Group},
  doi={10.1038/s41928-020-0410-3}
}

@article{xor_memristor_searchd,
  title={Searchd: A memory-centric hyperdimensional computing with stochastic training},
  author={Imani, Mohsen and Yin, Xunzhao and Messerly, John and Gupta, Saransh and Niemier, Michael and Hu, Xiaobo Sharon and Rosing, Tajana},
  journal={IEEE Transactions on Computer-Aided Design of Integrated Circuits and Systems},
  volume={39},
  number={10},
  pages={2422--2433},
  year={2019},
  publisher={IEEE},
  doi={ 10.1109/TCAD.2019.2952544}
}

@article{FeFET_IMC_HDC_XOR_majority,
  title={FeFET-based in-memory hyperdimensional encoding design},
  author={Huang, Qingrong and Yang, Zeyu and Ni, Kai and Imani, Mohsen and Zhuo, Cheng and Yin, Xunzhao},
  journal={IEEE Transactions on Computer-Aided Design of Integrated Circuits and Systems},
  volume={42},
  number={11},
  pages={3829--3839},
  year={2023},
  publisher={IEEE},
  doi={10.1109/TCAD.2023.3253766}
}

@inproceedings{dunkel2017fefet,
  title={A FeFET based super-low-power ultra-fast embedded NVM technology for 22nm FDSOI and beyond},
  author={D{\"u}nkel, Stefan and Trentzsch, Martin and Richter, Regina and Moll, Patrick and Fuchs, Christine and Gehring, Oliver and Majer, Mateusz and Wittek, Stefan and M{\"u}ller, B and Melde, Thomas and Mulaosmanovic, H. and Slesazeck, S. and Müller, S. and Ocker, J. and Noack, M. and Löhr, D.-A. and Polakowski, P.},
  booktitle={2017 IEEE International Electron Devices Meeting (IEDM)},
  pages={19--7},
  year={2017},
  organization={IEEE},
  doi={10.1109/IEDM.2017.8268425}
}

@article{Nature_FeFET_variability,
  title={First demonstration of in-memory computing crossbar using multi-level Cell FeFET},
  author={Soliman, Taha and Chatterjee, Swetaki and Laleni, Nellie and M{\"u}ller, Franz and Kirchner, Tobias and Wehn, Norbert and K{\"a}mpfe, Thomas and Chauhan, Yogesh Singh and Amrouch, Hussam},
  journal={Nature Communications},
  volume={14},
  number={1},
  pages={6348},
  year={2023},
  publisher={Nature Publishing Group UK London},
  doi={10.1038/s41467-023-42110-y}
}

@article{swetaki_temperature_variability,
  title={Temperature-and variability-aware compact modeling of ferroelectric FDSOI FET for memory and emerging applications},
  author={Chatterjee, Swetaki and Kumar, Shubham and Gaidhane, Amol and Dabhi, Chetan Kumar and Chauhan, Yogesh Singh and Amrouch, Hussam},
  journal={Solid-State Electronics},
  pages={108954},
  year={2024},
  publisher={Elsevier},
  doi={10.1016/j.sse.2024.108954}
}

@INPROCEEDINGS{Musaib_majority,
  author={Rafiq, Musaib and Chauhan, Yogesh Singh and Sahay, Shubham},
  booktitle={2024 8th IEEE Electron Devices Technology \& Manufacturing Conference (EDTM)}, 
  title={Exploiting Single Ferroelectric FET for Efficient Implementation of Majority Gate Function for Approximate Computing}, 
  year={2024},
  volume={},
  number={},
  pages={1-3},
  doi={10.1109/EDTM58488.2024.10511629}}

@article{mulaosmanovic2021ferroelectric,
  title={Ferroelectric field-effect transistors based on HfO2: a review},
  author={Mulaosmanovic, H. and Breyer, E. T. and D{\"u}nkel, S. and Beyer, S. and Mikolajick, T. and Slesazeck, S.},
  journal={Nanotechnology},
  volume={32},
  number={50},
  pages={pp. 502002},
  year={2021},
  month={9},
  publisher={IOP Publishing},
  doi={10.1088/1361-6528/ac189f}
}

@inproceedings{FeFETendurance10to11,
  title={Ultra-low power robust 3bit/cell Hf 0.5 Zr 0.5 O 2 ferroelectric FinFET with high endurance for advanced computing-in-memory technology},
  author={De, Sourav and Lu, Darsen D and Le, Hoang-Hiep and Mazumder, Soumen and Lee, Yao-Jen and Tseng, Wei-Chih and Qiu, Bo-Han and Baig, Md Aftab and Sung, Po-Jung and Su, Chung-Jun and Wu, Chien-Ting and Wu, Wen-Fa and Yeh, Wen-Kuan and Wang, Yeong-Her},
  booktitle={2021 symposium on VLSI technology},
  pages={1--2},
  year={2021},
  organization={IEEE},
  doi={}
}

@article{MORRAM_review_1,
  title={Metal--oxide RRAM},
  author={Wong, H-S Philip and Lee, Heng-Yuan and Yu, Shimeng and Chen, Yu-Sheng and Wu, Yi and Chen, Pang-Shiu and Lee, Byoungil and Chen, Frederick T and Tsai, Ming-Jinn},
  journal={Proceedings of the IEEE},
  volume={100},
  number={6},
  pages={1951--1970},
  month={6},
  year={2012},
  publisher={IEEE},
  doi={10.1109/JPROC.2012.2190369}
}

@article{RRAM_review_2,
  title={Resistive random access memory (RRAM) technology: From material, device, selector, 3D integration to bottom-up fabrication},
  author={Chen, Hong-Yu and Brivio, Stefano and Chang, Che-Chia and Frascaroli, Jacopo and Hou, Tuo-Hung and Hudec, Boris and Liu, Ming and Lv, Hangbing and Molas, Gabriel and Sohn, Joon and Spiga, Sabina and Mani Teja, V. and Vianello, Elisa and Philip Wong, H.-S.},
  journal={Journal of Electroceramics},
  volume={39},
  pages={21--38},
  month={6},
  year={2017},
  publisher={Springer},
  doi={10.1007/s10832-017-0095-9}
}

@INPROCEEDINGS{STTMRAM_3D,
  author={Huai, Yiming and Yang, Hongxin and Hao, Xiaojie and Wang, Zihui and Malmhall, Roger and Sato, Kimihiro and Zhang, Jing and Jung, Dong Ha and Wang, Xiaobin and Xu, Pengfa and Yen, Bing K.},
  booktitle={2018 IEEE International Memory Workshop (IMW)}, 
  title={High Density 3D Cross-Point STT-MRAM}, 
  month={5},
  year={2018},
  volume={},
  number={},
  pages={1-4},
  doi={10.1109/IMW.2018.8388833}}

@INPROCEEDINGS{MTJ_endurance_retention,
  author={Sato, H. and Honjo, H. and Watanabe, T. and Niwa, M. and Koike, H. and Miura, S. and Saito, T. and Inoue, H. and Nasuno, T. and Tanigawa, T. and Noguchi, Y. and Yoshiduka, T. and Yasuhira, M. and Ikeda, S. and Kang, S.- Y. and Kubo, T. and Yamashita, K. and Yagi, Y. and Tamura, R. and Endoh, T.},
  booktitle={2018 IEEE International Electron Devices Meeting (IEDM)}, 
  title={14ns write speed 128Mb density Embedded STT-MRAM with endurance$>$1010 and 10yrs retention@85°C using novel low damage MTJ integration process}, 
  month={12},
  year={2018},
  volume={},
  number={},
  pages={27.2.1-27.2.4},
  doi={10.1109/IEDM.2018.8614606}}

@article{PCM_review_1,
  title={Phase-change memory—Towards a storage-class memory},
  author={Fong, Scott W and Neumann, Christopher M and Wong, H-S Philip},
  journal={IEEE Transactions on Electron Devices},
  volume={64},
  number={11},
  pages={4374--4385},
  month={11},
  year={2017},
  publisher={IEEE},
  doi={10.1109/TED.2017.2746342}
}

@article{PCM_review_2,
  title={Enhancing the performance of phase change memory for embedded applications},
  author={Li, Xi and Chen, Houpeng and Xie, Chenchen and Cai, Daolin and Song, Sannian and Chen, Yifeng and Lei, Yu and Zhu, Min and Song, ZhiTang},
  journal={physica status solidi (RRL)--Rapid Research Letters},
  volume={13},
  number={4},
  pages={1800558},
  month={2},
  year={2019},
  publisher={Wiley Online Library},
  doi={10.1002/pssr.201800558}
}

@article{kanerva2009hyperdimensional_review,
  title={Hyperdimensional computing: An introduction to computing in distributed representation with high-dimensional random vectors},
  author={Kanerva, Pentti},
  journal={Cognitive computation},
  volume={1},
  number={2},
  pages={139--159},
  year={2009},
  publisher={Springer},
  doi={10.1007/s12559-009-9009-8}
}

@article{HDC_review,
  title={Classification using hyperdimensional computing: A review},
  author={Ge, Lulu and Parhi, Keshab K},
  journal={IEEE Circuits and Systems Magazine},
  volume={20},
  number={2},
  pages={30--47},
  year={2020},
  publisher={IEEE},
  doi={10.1109/MCAS.2020.2988388}
}

@inproceedings{FeFET_switching_chargetrapping_mulaosmanovic,
  title={Switching and charge trapping in HfO 2-based ferroelectric FETs: An overview and potential applications},
  author={Mulaosmanovic, Halid and Breyer, Evelyn T and Mikolajick, Thomas and Slesazeck, Stefan},
  booktitle={2020 4th IEEE Electron Devices Technology \& Manufacturing Conference (EDTM)},
  pages={1--4},
  year={2020},
  organization={IEEE},
  doi={10.1109/EDTM47692.2020.9118005}
}

@inproceedings{Record-based_better_tha_N-gram_imani2018hierarchical,
  title={Hierarchical hyperdimensional computing for energy efficient classification},
  author={Imani, Mohsen and Huang, Chenyu and Kong, Deqian and Rosing, Tajana},
  booktitle={Proceedings of the 55th Annual Design Automation Conference},
  pages={1--6},
  year={2018},
  doi={10.1145/3195970.3196060}

}

\end{document}